\documentclass[11pt, a4paper]{article}

\usepackage[T1]{fontenc}
\usepackage[utf8]{inputenc}
\usepackage{lmodern}
\usepackage[margin=2.5cm]{geometry}
\usepackage{microtype}
\usepackage{graphicx}
\usepackage{booktabs}
\usepackage{multirow}
\usepackage{amsmath}
\usepackage{amssymb}
\usepackage{xcolor}
\usepackage{hyperref}
\usepackage{enumitem}
\usepackage{caption}
\usepackage{subcaption}
\usepackage{listings}
\usepackage{fancyhdr}
\usepackage{titlesec}
\usepackage{abstract}
\usepackage{parskip}
\usepackage{float}
\usepackage{array}
\usepackage{tabularx}
\usepackage{colortbl}
\usepackage{tikz}
\usetikzlibrary{arrows.meta, positioning, shapes.geometric}

\usepackage{newunicodechar}
\newunicodechar{—}{\textemdash}
\newunicodechar{–}{\textendash}
\newunicodechar{→}{\ensuremath{\rightarrow}}
\newunicodechar{★}{\ensuremath{\star}}
\newunicodechar{×}{\ensuremath{\times}}
\newunicodechar{≥}{\ensuremath{\geq}}
\newunicodechar{≤}{\ensuremath{\leq}}
\newunicodechar{≈}{\ensuremath{\approx}}
\newunicodechar{∼}{\ensuremath{\sim}}

\definecolor{monavecblue}{HTML}{2563EB}
\definecolor{darkgray}{HTML}{374151}
\definecolor{lightgray}{HTML}{F3F4F6}
\definecolor{accent}{HTML}{16A34A}
\definecolor{warnorange}{HTML}{EA580C}
\definecolor{codebg}{HTML}{F8F8F8}

\hypersetup{
    colorlinks=true,
    linkcolor=monavecblue,
    citecolor=monavecblue,
    urlcolor=monavecblue,
    pdftitle={MonaVec: A Training-Free Embedded Vector Search Kernel for Edge AI},
    pdfauthor={Oguzhan Yenen},
}

\titleformat{\section}{\large\bfseries\color{darkgray}}{{\thesection}}{0.8em}{}[\vspace{-0.3em}\rule{\linewidth}{0.4pt}]
\titleformat{\subsection}{\normalsize\bfseries\color{darkgray}}{\thesubsection}{0.8em}{}
\titleformat{\subsubsection}{\normalsize\itshape}{\thesubsubsection}{0.8em}{}

\newcommand{\monavec}{\textsc{MonaVec}}
\newcommand{\mvec}{\texttt{.mvec}}
\newcommand{\rhdh}{\textsc{RHDH}}
\newcommand{\norm}[1]{\left\lVert#1\right\rVert}

\begin{document}

\begin{center}
    {\LARGE\bfseries\color{darkgray}
        \monavec: A Training-Free Embedded Vector Search Kernel\\[6pt]
        for Edge and Offline AI Systems
    }\\[12pt]
    {\large O\u{g}uzhan Yenen}\\[4pt]
    {\normalsize Mona \quad \texttt{oguzhanyenen@gmail.com}}\\[6pt]
    {\small Preprint — \today}
\end{center}

\vspace{0.5em}
\noindent\rule{\linewidth}{0.8pt}
\vspace{0.3em}

\begin{abstract}
\noindent
We present \monavec{}, a deterministic, embedded vector-search kernel for edge
and offline AI---settings where server infrastructure, network connectivity, and
training data are all unavailable. Existing vector-search systems assume a
persistent server, gigabytes of RAM, or a training pass over the corpus;
\monavec{} instead targets the deployment profile of SQLite: one file, one
function call, runs anywhere.

Its quantization core is \textbf{training-free by default} and
\emph{data-oblivious}: a Randomized Hadamard Transform (\rhdh{}) conditions any
input distribution toward $\mathcal{N}(0,1)$, so precomputed Lloyd-Max tables
quantize to 4 bits ($8\times$ smaller) with no learned codebook and no data pass.

The index persists as a single \mvec{} file whose embedded ChaCha20 rotation
seed makes results \emph{reproducible across architectures} and
\emph{byte-identical within a build}---a determinism guarantee that
parallel-build graph libraries cannot offer.

On semantic embeddings (AG News, 45K\,$\times$\,1024-dim BGE-M3, cosine),
\monavec{} 4-bit BruteForce reaches \textbf{0.960 Recall@10 in 27\,MB}---leading
float32 FAISS-IVF and 8-bit usearch on recall---while trading peak throughput for
byte-identical determinism. A single-pass global standardization (\texttt{fit()})
extends the same data-oblivious pipeline to magnitude-sensitive L2 data, and
optional IvfFlat and HNSW backends carry it to million-vector corpora.

\monavec{} is implemented in pure Rust with Python bindings and runtime SIMD
dispatch (AVX-512/AVX2/NEON/scalar). It targets on-device RAG, offline agents,
and embedded retrieval---the niche SQLite occupies for relational data: one file,
one call, runs anywhere.
\end{abstract}
\vspace{0.5em}
\noindent\rule{\linewidth}{0.4pt}

\section{Introduction}

Large language model inference is rapidly migrating to edge devices.
Frameworks such as \texttt{llama.cpp}~\cite{llama_cpp} and MLC-LLM~\cite{mlc_llm}
now enable billion-parameter models to run on mobile and embedded hardware with
sub-100ms token generation latency.
However, retrieval-augmented generation (RAG)~\cite{lewis2020rag} and semantic
search—critical components of production AI pipelines—remain server-bound.
Existing vector search systems such as Qdrant~\cite{qdrant} and
Weaviate~\cite{weaviate} require persistent server processes and gigabytes of RAM\@.
FAISS~\cite{johnson2019faiss}, while embeddable, requires training data for IVF
clustering and exposes a C++ API that is unwieldy for embedded or mobile contexts.

This gap motivates \monavec{}: an embedded vector search kernel with the deployment
profile of SQLite~\cite{sqlite} applied to vector search.
\emph{Zero server. Zero network. Zero training data.}
A single file, loadable with a single function call—on a Raspberry Pi,
a mobile device, an offline laptop, or a production server.
We use \emph{training-free} for \monavec{}'s default configuration, whose
quantizer is parameterized without observing any data; Table~\ref{tab:data_dependence}
gives a precise taxonomy of the few optional components---an L2 calibration pass,
graph construction, and an opt-in trained partitioner---that do read the corpus.

\monavec{} is more than a library kernel. It ships as a complete end-to-end system:
a Rust crate (\texttt{monavec-core}), Python bindings (\texttt{monavec}),
a FastAPI service layer with REST API, a web-based admin UI, a CLI,
hybrid sparse-dense retrieval with BM25, and pluggable identity-based multi-tenancy.
The entire stack is designed to work offline, with no external services required.

\vspace{0.5em}
\noindent\textbf{Contributions.} This paper makes the following contributions:

\begin{enumerate}[leftmargin=1.5em, itemsep=2pt]
    \item A \textbf{data-oblivious quantization pipeline} combining \rhdh{} rotation
          with Lloyd-Max optimal scalar quantization~\cite{lloyd1982, max1960}.
          After random rotation, all coordinates approximate $\mathcal{N}(0,1)$
          in high dimension---the concentrated marginal of a randomly rotated unit
          vector~\cite{vempala2004}, realized by the fast \rhdh{}~\cite{ailon2009}.
          This allows precomputed optimal centroid tables without any training pass.

    \item \textbf{Global scalar standardization for L2 metric} (\texttt{fit()}).
          For raw-magnitude vectors (pixels, SIFT descriptors), per-dimension
          normalization would destroy Euclidean distance ordering.
          We introduce a single-pass global standardization—applying a scalar
          $(x - \mu)/\sigma$ uniformly across all dimensions—that restores the
          $\mathcal{N}(0,1)$ quantizer assumption while preserving L2 ordering.
          On fashion-mnist-784, Recall@10 improves from 0.41 to 0.62 (+52\%).

    \item \textbf{Metric-aware HNSW graph construction.}
          We show that building the HNSW graph with dot product scoring
          (correct for Cosine) but searching with L2 scoring
          yields corrupt graph topology for L2 metrics, since the greedy
          traversal navigates toward the wrong neighbours.
          Replacing graph-construction scoring with
          $\langle q, v \rangle - \frac{1}{2}\norm{v}^2$ for L2 recovers
          correct topology and improves Recall@10 from 0.31 to 0.62.

    \item \textbf{Auto-M policy for HNSW at scale.}
          We empirically show that the standard M=32 parameter is insufficient
          for N\,$\geq$\,1M vectors: at N=1.18M, M=32 yields Recall@10=0.800
          while M=64 achieves 0.850 at identical QPS.
          Graph diameter grows with N; higher M compensates by reducing the
          effective diameter. We adopt an automatic policy: M=32 for N\,<\,10$^6$,
          M=64 for N\,$\geq$\,10$^6$.

    \item \textbf{FP32-build / 4-bit-search HNSW}: graph topology constructed
          with exact dot products to avoid corruption from quantization noise,
          while storage and query scoring use 4-bit packed vectors.

    \item \textbf{Pre-filter allowlist}: applied before scoring (not post-filter),
          preserving recall in filtered retrieval regardless of allowlist selectivity.

    \item \textbf{Runtime SIMD dispatch}: AVX-512F+BW, AVX2+FMA, NEON, and scalar
          paths compiled into a single binary; selected at runtime.

    \item \textbf{Hybrid dense+sparse retrieval}: BM25 sparse index co-located
          with dense vector index. Results fused via Reciprocal Rank Fusion (RRF),
          enabling keyword-aware semantic search with no external dependencies.

    \item \textbf{Pluggable identity-based multi-tenancy}: token-based namespace
          isolation via a single HTTP endpoint contract, compatible with any
          OAuth2/JWT/LDAP system or standalone token-as-namespace mode.
\end{enumerate}

\section{Design Philosophy and Language Choice}

\subsection{Why Rust}

\monavec{} is implemented in pure Rust with zero C dependencies.
This choice is motivated by three constraints inherent to the edge deployment target:

\paragraph{Determinism.}
Edge AI systems—medical devices, offline agents, mobile applications—cannot
tolerate non-deterministic behavior. A query on a re-loaded index must return
identical results. Rust's ownership model prevents data races; ChaCha20 seeding
of the RHDH rotation matrix is stored in the \mvec{} file, so reloading and
searching the same index reproduces the same top-K results on any platform---and
is byte-identical within a given build.
This is a property we call \emph{portable determinism}: the same \mvec{} file
produces the same top-K results on an x86 server, an ARM laptop, and a RISC-V
embedded device. Determinism requires care with floating-point evaluation order,
since IEEE-754 arithmetic~\cite{ieee754} is not associative and reordering a
reduction changes the rounded result~\cite{goldberg1991fp}. Within a given build,
\monavec{} fixes the seed, the SIMD reduction order, and a single
micro-architecture baseline (Section~\ref{sec:scoring-kernel}, \S3.7), so the same
inputs yield the same bits. We use three notions precisely. \emph{Deterministic}:
a build reproduces its own results exactly. \emph{Byte-identical}: the score
vector is equal bit-for-bit---a guarantee \monavec{} provides \emph{within a
build}. \emph{Reproducible}: the same \mvec{} file yields the same top-K on any
machine---what \monavec{} guarantees \emph{across} architectures, since the
heterogeneous SIMD kernels (AVX-512/AVX2/NEON/scalar) agree only to within their
validated tolerance ($10^{-4}$, \S3.7) rather than bit-for-bit. The deterministic
\mvec{} encoding---an integer-seeded rotation and table-lookup quantization that
produce the same packed bytes on any platform---is what makes both properties
possible.

A direct consequence is that index construction is \emph{sequential and
single-threaded by design}. Parallel graph construction—as used by
hnswlib and usearch—makes insertion order non-deterministic, and therefore the
resulting graph topology non-deterministic: the same vectors yield a different
index on each run. MonaVec deliberately forgoes this parallelism to preserve
reproducibility. The cost is slower build time relative to parallel
implementations (Section~\ref{sec:competitors}); we consider this an acceptable
trade for the target domains (medical, offline, embedded) where a non-reproducible
index is a liability. Closing the build-time gap \emph{without} sacrificing
determinism—e.g.\ a hand-written AVX2 build-distance kernel—is future work.

\paragraph{Zero C dependencies.}
FAISS requires BLAS, LAPACK, and optionally CUDA—external C/C++ libraries
that are unavailable or impractical to compile on constrained targets.
\monavec{}'s core has no C FFI, no BLAS calls, no OpenMP. SIMD acceleration
is provided via the \texttt{wide} crate (portable AVX2/NEON) and manual
intrinsics for the nibble-scoring hot path only.
\texttt{cargo build --target aarch64-unknown-linux-gnu} produces a working binary
with no additional system dependencies.

\paragraph{Memory safety without runtime overhead.}
Rust's compile-time memory safety eliminates buffer overflows and use-after-free
errors—a critical property for a library that will be embedded in user applications.
Unlike garbage-collected runtimes, Rust imposes zero runtime overhead,
keeping latency deterministic and predictable.

\subsection{Target Deployment Profile}

\monavec{} is designed to occupy the same deployment niche as SQLite: a library
that applications link against, not a service that applications connect to.
The primary targets are:

\begin{itemize}[itemsep=2pt]
    \item \textbf{Mobile devices}: iOS, Android apps with on-device RAG
    \item \textbf{Edge AI hardware}: Raspberry Pi, NVIDIA Jetson, embedded Linux
    \item \textbf{Offline agents}: LLM agents in air-gapped environments
    \item \textbf{Desktop applications}: offline document search, local knowledge bases
    \item \textbf{Production servers}: as a lightweight alternative where server infrastructure is undesirable
\end{itemize}

The service layer (FastAPI REST API, CLI, Docker) enables \monavec{} to scale
from an embedded library to a production microservice without code changes—
the same \mvec{} files, the same Python API, the same results.

\section{System Design}

\subsection{Quantization Pipeline}

\begin{figure}[H]
    \centering
    \includegraphics[width=\linewidth]{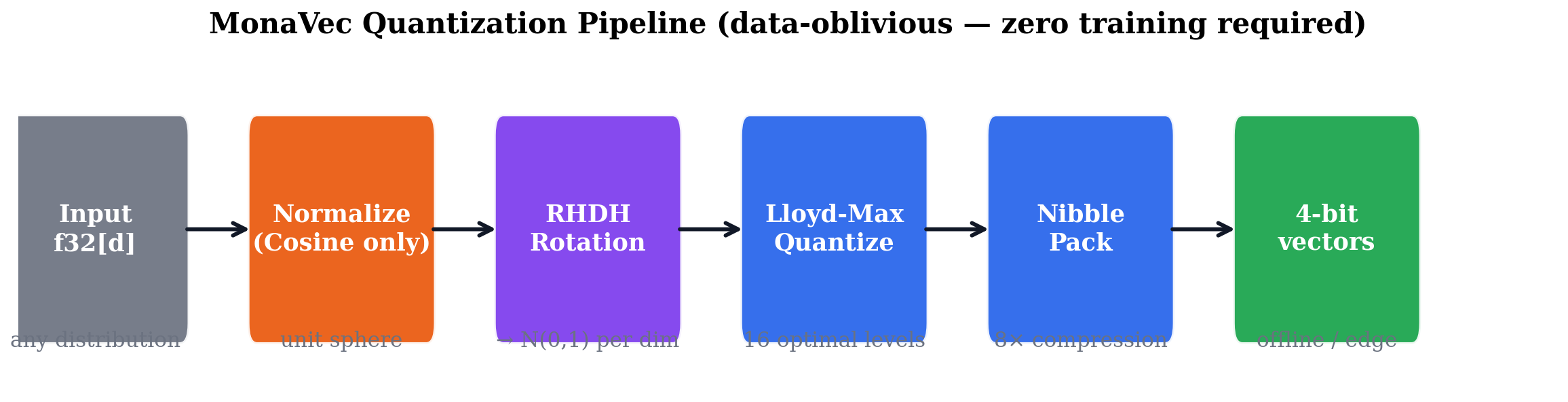}
    \caption{
        \monavec{} quantization pipeline. Cosine inputs are unit-normalized;
        L2 inputs optionally standardized via \texttt{fit()}; Dot inputs are raw.
        All paths share the RHDH rotation and Lloyd-Max quantization stages.
    }
    \label{fig:pipeline}
\end{figure}

The full quantization pipeline is entirely data-oblivious for Cosine and Dot metrics.
For L2, a single-pass calibration step (\texttt{fit()}) is optionally applied
before encoding. Figure~\ref{fig:pipeline} illustrates both paths.

\paragraph{A taxonomy of data dependence.}
Because \monavec{} is described as ``training-free,'' it is worth stating
precisely which stages touch the data. The default configuration---BruteForce
search over the \rhdh{}\,+\,Lloyd-Max pipeline---is \emph{data-oblivious}: no
stage observes the corpus when the quantizer is parameterized. The components that
\emph{do} read data are optional, and each falls into a distinct category
(Table~\ref{tab:data_dependence}): a single-pass, non-iterative \emph{calibration}
(the L2 standardization \texttt{fit()}, \S3.1); deterministic \emph{index
construction} that learns no codebook (the HNSW graph, \S3.4); and an explicitly
opt-in \emph{trained} partitioner (IvfFlat $k$-means, \S3.4). Only the last is
training in the usual sense---iterative optimization over a representative set. We
use \emph{training-free} throughout to denote the default; the opt-in IvfFlat
index is its single exception.

\begin{table}[H]
\centering
\caption{Data dependence of each \monavec{} component. The default configuration
(BruteForce over \rhdh{}\,+\,Lloyd-Max) is data-oblivious end to end; the optional
components are classified by \emph{how} they read the corpus, not merely
\emph{whether} they do.}
\label{tab:data_dependence}
\footnotesize
\begin{tabular}{lll}
\toprule
\textbf{Component} & \textbf{Category} & \textbf{Reads the corpus?} \\
\midrule
\rhdh{} rotation (ChaCha20 seed)        & Data-oblivious     & No (seed only) \\
Lloyd-Max $\mathcal{N}(0,1)$ tables     & Data-oblivious     & No (precomputed offline) \\
L2 standardization (\texttt{fit()})     & Calibration        & One pass, summary statistics \\
HNSW graph construction                 & Index construction & Geometry only, no codebook \\
IvfFlat partitioning ($k$-means)        & Training (opt-in)  & Iterative over the corpus \\
\bottomrule
\end{tabular}
\end{table}

\subsubsection{Step 1 — Metric-Aware Input Preparation}

\paragraph{Cosine similarity.}
Inputs are normalized to unit length:
\[
    \hat{v} = \frac{v}{\norm{v}_2}
\]
This makes dot product in the quantized space equivalent to cosine similarity
in the original space, and ensures the RHDH output follows $\mathcal{N}(0,1)$
(since all input vectors lie on the unit sphere).

\paragraph{L2 (Euclidean distance).}
L2 distance is magnitude-sensitive: $\norm{a - b}_2$ depends on the absolute
values of both vectors. Applying unit normalization would collapse all points to
the unit sphere, making L2 equivalent to Cosine and destroying the geometric
information the metric is intended to capture.

Instead, \monavec{} applies \emph{global scalar standardization} when
\texttt{fit(sample)} is called:
\[
    x_{\text{std}} = \frac{x - \mu_{\text{global}}}{\sigma_{\text{global}}}
\]
where $\mu_{\text{global}}$ and $\sigma_{\text{global}}$ are scalar statistics
computed once over a representative sample—a single pass, no iteration.
The key property: applying the \emph{same scalar} to every dimension is a
uniform scaling, which preserves relative L2 distances:
\[
    \norm{a_{\text{std}} - b_{\text{std}}}_2 = \frac{\norm{a - b}_2}{\sigma_{\text{global}}}
\]
Ranking by $\norm{a_{\text{std}} - b_{\text{std}}}_2$ is equivalent to
ranking by $\norm{a - b}_2$. Only the scale changes, not the ordering.

\paragraph{Why not per-dimension standardization?}
Per-dimension whitening applies a different scale to each dimension:
$x_i \leftarrow (x_i - \mu_i) / \sigma_i$.
This changes the metric from Euclidean to Mahalanobis—nearest neighbours
in whitened space are not the same as nearest neighbours in original space.
On fashion-mnist, per-dimension whitening achieves Recall@10 = 0.53
while global standardization achieves 0.62. The gap confirms that preserving
L2 ordering is more important than per-dimension distribution matching.

\paragraph{Dot product.}
Raw vectors are passed through unchanged. Magnitude is a signal—intentional.

\subsubsection{Step 2 — Randomized Hadamard Transform (RHDH)}

The \rhdh{} is a structured random rotation:
\[
    R = \frac{1}{\sqrt{d'}} H D
\]
where $H \in \{-1,+1\}^{d' \times d'}$ is a Walsh-Hadamard matrix,
$D = \text{diag}(r_1, \ldots, r_{d'})$ with $r_i \stackrel{\text{iid}}{\sim}
\text{Uniform}\{-1, +1\}$, $d'$ is the smallest power of 2 $\geq d$.
Random signs are generated from a ChaCha20~\cite{bernstein2008chacha} stream
seeded from the \mvec{} header—ensuring reproducible rotation across
all platforms and sessions.

\paragraph{Why \rhdh{} produces approximately $\mathcal{N}(0,1)$.}
After normalization and Hadamard rotation, each output coordinate is a
normalized sum of $d'$ sign-flipped terms. A coordinate of a randomly rotated
unit vector follows a concentrated Beta marginal that converges to
$\mathcal{N}(0, 1/d')$ as $d'$ grows~\cite{vempala2004}; the structured \rhdh{} is
the fast $O(d\log d)$ surrogate~\cite{ailon2009} for such a rotation. After scaling
by $\sqrt{d'}$, coordinates are approximately $\mathcal{N}(0,1)$ at the embedding
dimensionalities we target. This near-distributional guarantee is the foundation for
training-free quantization: the distribution of inputs to the quantizer
is known \emph{in advance}, without observing any data.

The transform runs in $O(d \log d)$—substantially cheaper than the
$O(d^2)$ application of a full random orthogonal matrix.

\paragraph{Relation to rotation-based quantization.}
Conditioning a distribution by a random (or learned) rotation before low-bit
quantization is an established technique. QuaRot~\cite{ashkboos2024quarot} and
SpinQuant~\cite{liu2024spinquant} apply Hadamard-style rotations to remove
activation outliers for 4-bit LLM inference, and TurboQuant~\cite{turboquant}
develops it into a general-purpose vector quantizer
(Section~\ref{sec:turboquant}). \monavec{}'s contribution is not the rotation
primitive itself but its adaptation to the embedded-retrieval setting: the
rotation makes precomputed $\mathcal{N}(0,1)$ Lloyd-Max tables valid without any
data pass, and is co-designed with the index backends, asymmetric scoring, and a
deterministic single-file format (Section~\ref{sec:turboquant}).

\subsubsection{Step 3 — Lloyd-Max Scalar Quantization}

Given that \rhdh{} outputs are $\mathcal{N}(0,1)$-distributed, we apply
Lloyd-Max quantization~\cite{lloyd1982, max1960} with tables precomputed
for $\mathcal{N}(0,1)$.

\begin{figure}[H]
    \centering
    \includegraphics[width=0.82\linewidth]{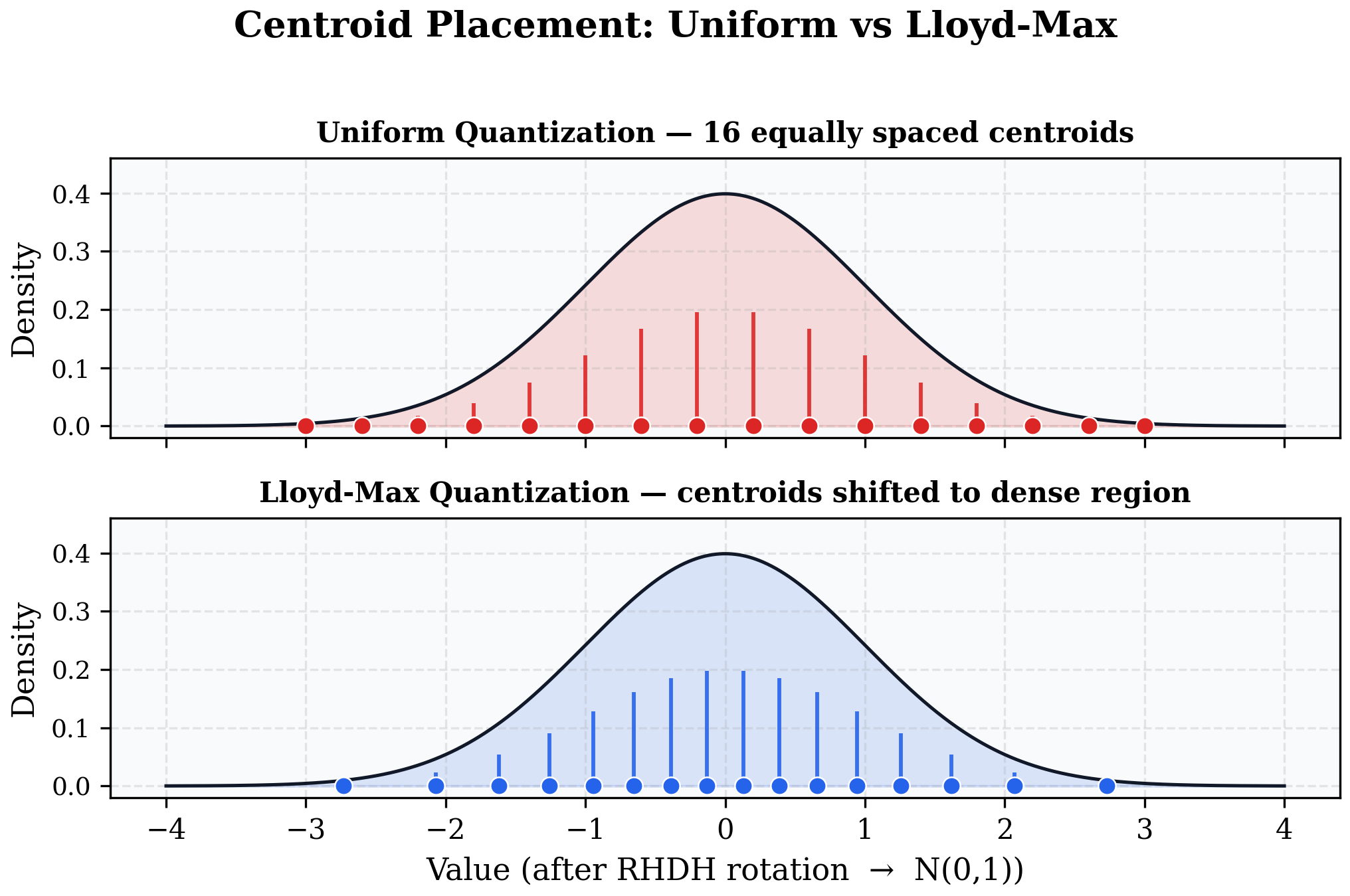}
    \caption{
        Centroid placement for uniform quantization versus Lloyd-Max on
        $\mathcal{N}(0,1)$ data. Lloyd-Max concentrates centroids where the
        density is highest—near zero—minimizing the mean squared error that
        dominates quantization distortion.
    }
    \label{fig:lloyd_max}
\end{figure}

\paragraph{Why Lloyd-Max over uniform quantization.}
$\mathcal{N}(0,1)$ data is dense near zero and sparse at the tails.
A uniform quantizer wastes precision on rarely-visited extreme values.
Lloyd-Max minimizes expected squared error by placing centroids where
data is dense—near zero—and spacing them more widely at the tails
(Figure~\ref{fig:lloyd_max}).
This yields 2--5\% Recall@10 improvement at identical memory cost
(Table~\ref{tab:lloyd_ablation}).

\paragraph{Precomputed tables.}
The Lloyd-Max centroids and decision boundaries for $\mathcal{N}(0,1)$
are computed offline (2000 iterations to convergence, tolerance $10^{-12}$)
and compiled into the binary as constants:
\begin{itemize}[itemsep=1pt]
    \item 4-bit: 16 centroids, 15 decision boundaries
    \item 2-bit: 4 centroids, 3 decision boundaries
\end{itemize}
No runtime computation. No storage in the \mvec{} file.

\subsubsection{Step 4 — Nibble Packing}

Quantized 4-bit indices (0--15) are packed two per byte.
The packed representation achieves 8$\times$ compression over float32.
A separate norms array stores the per-vector quantized L2 norm
$q_{\text{norm}} = \norm{\hat{v}_q}_2$ for length-renormalized scoring.

\subsection{Mixed-Precision Bit Allocation}

After \rhdh{} rotation, dimensions carry unequal information.
High-variance dimensions are spread widely in the rotated space—they require
fine-grained precision for accurate dot product computation.
Low-variance dimensions cluster tightly—coarser quantization suffices.

\monavec{} profiles per-dimension variance across a sample of indexed
vectors and applies water-filling~\cite{cover2006elements} to allocate
the available bit budget optimally:
dimensions above the variance threshold receive 4-bit precision,
those below receive 2-bit. The threshold is derived analytically from
the desired average bit-width (e.g., 3-bit average).

Packed layout per vector: \texttt{[4-bit block | 2-bit block]},
allowing dimension count to be stored in the header and scoring to
dispatch to the correct kernel per block.

\begin{figure}[H]
    \centering
    \includegraphics[width=0.85\linewidth]{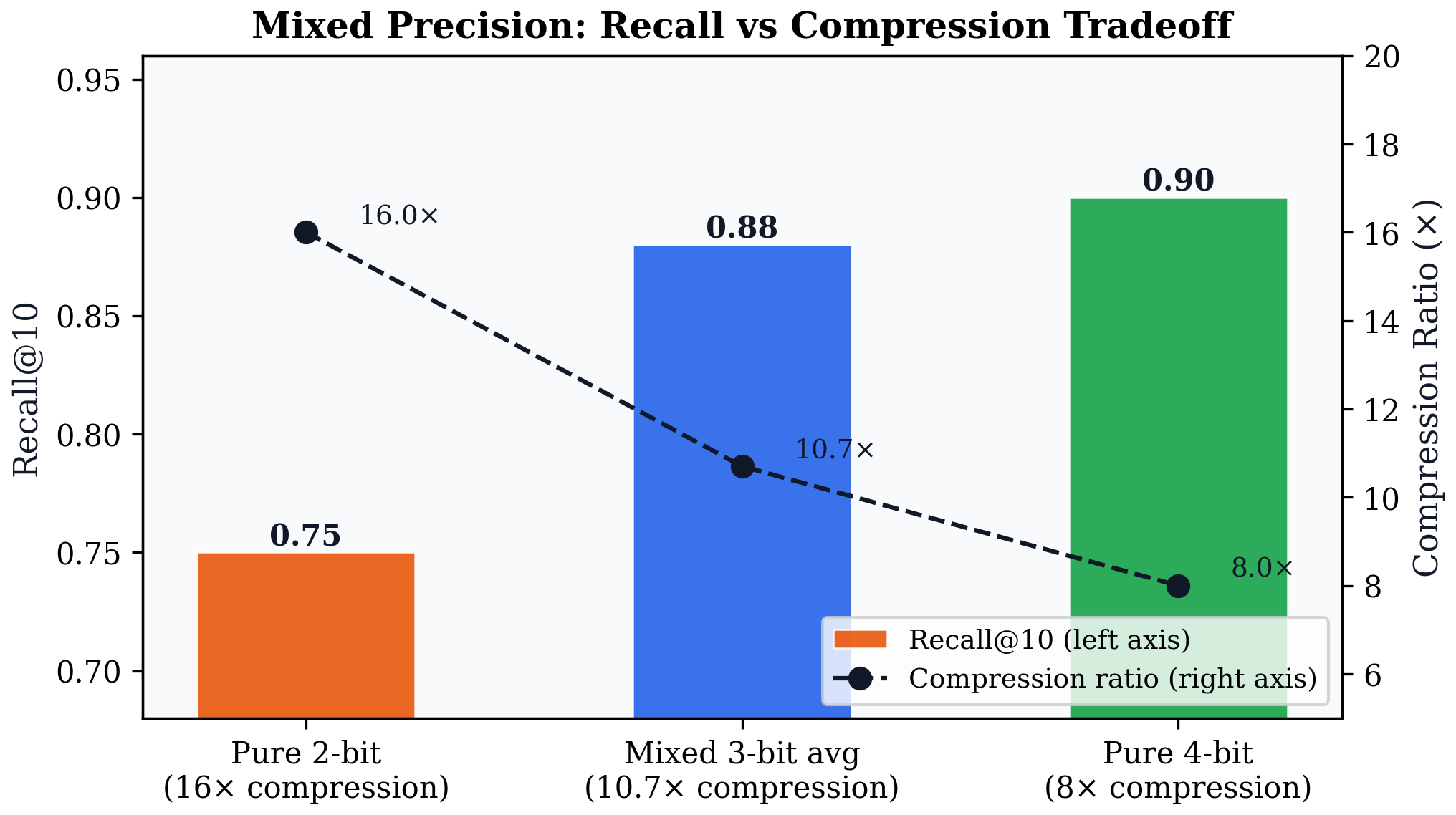}
    \caption{
        Mixed-precision quantization: Recall@10 and compression ratio for
        pure 2-bit, mixed 3-bit (average), and pure 4-bit configurations.
        Mixed precision improves the recall-per-byte tradeoff.
    }
    \label{fig:mixed}
\end{figure}

\paragraph{Implementation status.}
In the current implementation the 4-bit block holds the \emph{leading} dimensions
of the rotated vector: the variance-ordered permutation that would physically place
the highest-variance coordinates in the 4-bit block is computed but not yet persisted
in the file format. This matters less than it might appear, because RHDH equalizes
per-dimension variance by construction—the rotation is designed precisely to spread
information uniformly across coordinates. The residual variance structure that
water-filling exploits is therefore largest on data with pre-existing low-rank
structure (e.g.\ the synthetic Gaussian setting of Figure~\ref{fig:mixed}); a
format revision that stores the permutation, together with a systematic evaluation
on the real benchmark datasets, is future work (Section~\ref{sec:limitations}).

As shown in Figure~\ref{fig:mixed}, mixed 3-bit (average) achieves
10.7$\times$ compression with 0.88 Recall@10 on Gaussian test data,
compared to pure 4-bit (8$\times$ compression, 0.90 recall)—a Pareto
improvement when operating under strict memory budgets.
We note this result is established on synthetic $\mathcal{N}(0,1)$ data,
where the per-dimension variance structure that water-filling exploits is
fixed by construction. The head-to-head evaluations in
Section~\ref{sec:competitors} use uniform 4-bit throughout; a systematic
mixed-precision comparison on the real benchmark datasets and against
external baselines is left to future work (Section~\ref{sec:limitations}).

\subsection{Asymmetric Scoring and Length Renormalization}
\label{sec:asymmetric}

During search, query vectors remain in float32 while database vectors
are stored as packed 4-bit indices. This asymmetric scheme is critical:
if both query and database vectors were quantized, errors would accumulate
on both sides of the dot product. With only the database quantized,
the query retains full float32 precision.

Inspired by RaBitQ~\cite{gao2024rabitq}, \monavec{} stores a per-vector
quantized norm $q_{\text{norm}}$ and adjusts scores per metric:
\[
    s_{\text{Cosine}} = \frac{s_{\text{raw}}}{q_{\text{norm}}}
    \qquad
    s_{\text{Dot}} = s_{\text{raw}}
    \qquad
    s_{\text{L2}} = s_{\text{raw}} - \tfrac{1}{2} q_{\text{norm}}^2
\]
The L2 formula derives from $-\norm{q-v}^2 = 2\langle q, v\rangle - \norm{q}^2 - \norm{v}^2$:
dropping the query-constant $\norm{q}^2$ and approximating $\norm{v}^2$ by $q_{\text{norm}}^2$.

\subsection{Index Backends}

\monavec{} provides three index backends. All three share the same
quantization pipeline; they differ in how vectors are organized for retrieval.
The choice was driven by covering the full corpus-size spectrum
from edge (zero build cost) through production (sub-millisecond at scale).

\subsubsection{BruteForce}

Linear scan over all packed vectors. $O(n)$ per query, fully vectorized
with SIMD and memory-compact. Suitable for $n \lesssim 500$K on edge hardware.
Zero build time. Deterministic results. The recommended default for
embedded and offline deployments where simplicity and predictability
matter more than asymptotic complexity.

\textbf{Why include brute-force?} On ARM edge hardware (e.g.\ a Raspberry Pi
with NEON SIMD), a 4-bit brute-force scan of a moderate corpus completes within
the latency budget of many interactive edge use cases. For corpora that fit in
RAM, brute-force with SIMD often outperforms approximate methods that pay
graph-traversal overhead.

\subsubsection{IvfFlat}

Following the inverted-file (IVF) design~\cite{jegou2011pq}, \monavec{}
partitions vectors into $n_{\text{list}}$ clusters using Lloyd's algorithm.
At query time, the $n_{\text{probe}}$ nearest cluster centroids are identified
and their inverted lists scanned. Expected $O(n / n_{\text{list}} \times n_{\text{probe}})$
per query.

\textbf{Metric-aware k-means.}
IVF centroid computation is metric-sensitive:
\begin{itemize}[itemsep=1pt]
    \item \textbf{Cosine}: centroids are L2-normalized after each mean update
          (direction is the representative; magnitude is irrelevant).
    \item \textbf{Dot / L2}: centroids are raw means (magnitude preserved).
\end{itemize}

\textbf{Why IvfFlat over HNSW for some use cases?}
IvfFlat has lower memory overhead than HNSW (no graph edge storage),
predictable query time (probing a fixed number of cells), and simpler
tuning: one parameter ($n_{\text{probe}}$) controls the recall-speed tradeoff.
For N between 100K--1M with memory constraints, IvfFlat is often the
better choice.

\subsubsection{HNSW — FP32 Build, 4-bit Search, Auto-M}

HNSW~\cite{malkov2018hnsw} builds a multi-layer proximity graph where
each node connects to its $M$ nearest neighbors at its assigned layer.
Greedy traversal from entry point to query delivers $O(\log n)$ search.

\paragraph{FP32 build is necessary.}
Quantization noise magnitude is $\sim$0.01--0.02 on normalized embeddings.
The cosine score gap between true nearest neighbors is typically
$\sim$0.001--0.003. If graph construction uses 4-bit scores, quantization
noise exceeds the signal and incorrect neighbors are selected as edges—
corrupting the graph topology irreparably.
FP32 build preserves correct topology; 4-bit scoring during search introduces
only ranking noise, not structural damage.

\paragraph{Metric-aware graph construction.}
For Cosine/Dot metrics, greedy traversal during graph construction uses
plain dot product $\langle q, v \rangle$. For L2, this is incorrect:
two vectors may have high dot product while being far in Euclidean space.
\monavec{} uses
\[
    s_{\text{build}}^{L2}(q, v) = \langle q, v \rangle - \tfrac{1}{2}\norm{v}^2
\]
which correctly approximates $-\tfrac{1}{2}\norm{q-v}^2$ (up to the
query-constant $\norm{q}^2$). Without this fix, HNSW L2 achieves
Recall@10 = 0.31 on fashion-mnist; with the fix, 0.614 at ef=40 (0.624 at ef=400),
Table~\ref{tab:fashionmnist}.

\paragraph{M parameter and graph diameter.}
$M$ in HNSW controls the per-node degree—the number of bidirectional
connections each node maintains in the graph.
It is \emph{not} the search depth; that is controlled by \texttt{ef\_search}.
As corpus size $N$ grows, the graph diameter increases: the maximum shortest
path between any two nodes grows with $N$ when $M$ is fixed.
Greedy search relies on the graph being well-connected enough that
the true nearest neighbour is reachable from any starting point in
few hops. When diameter is too large (small $M$, large $N$), greedy search
terminates at a suboptimal local minimum.

\begin{figure}[H]
    \centering
    \includegraphics[width=\linewidth]{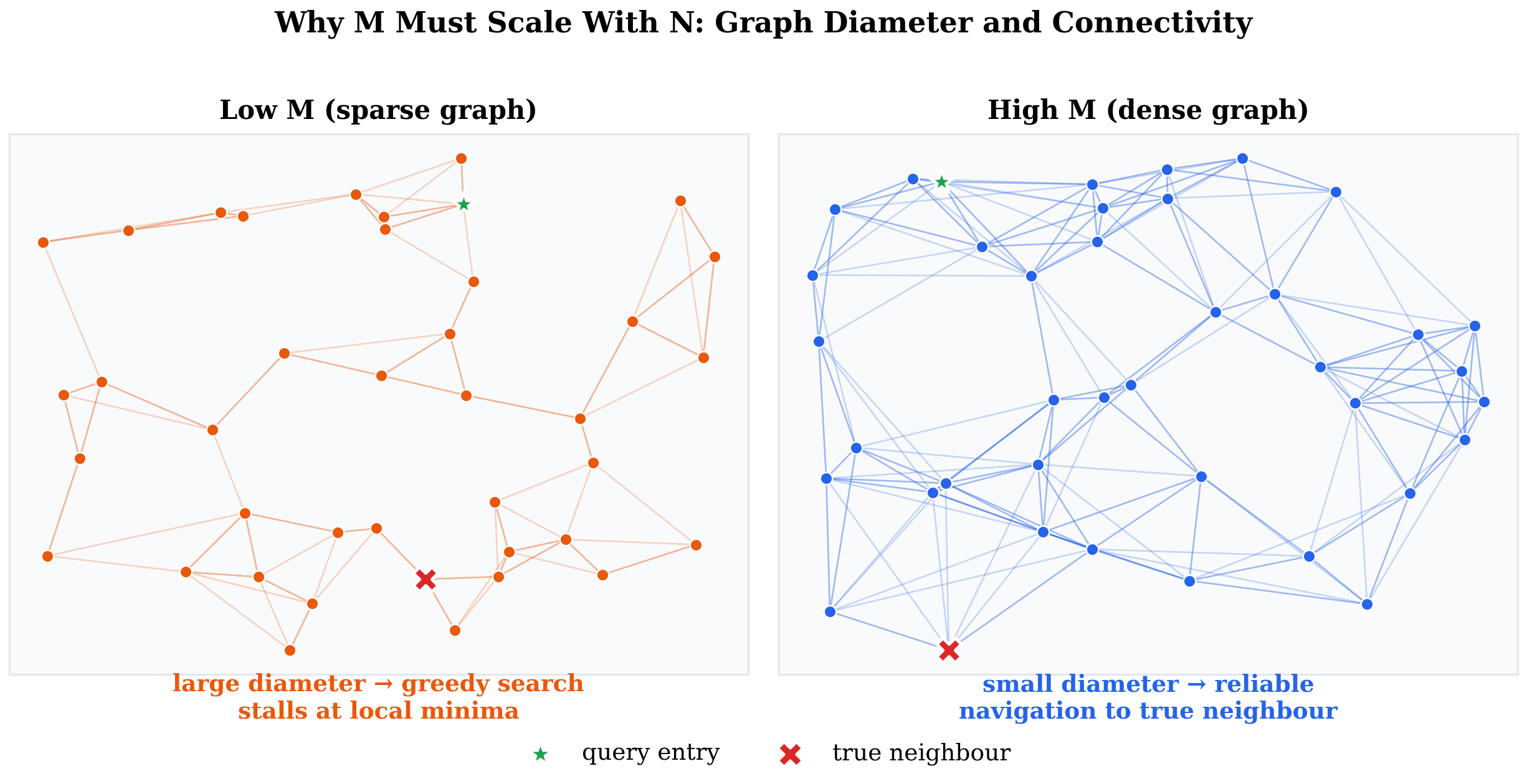}
    \caption{
        Why $M$ must scale with $N$. With low $M$ (left), the graph is sparse and
        its diameter is large—greedy search from the query entry ($\star$) frequently
        stalls before reaching the true neighbour ($\times$). With high $M$ (right),
        the dense graph has small diameter and reliable navigation. At N=1.18M, this
        is the difference between Recall@10 of 0.800 (M=32) and 0.850 (M=64).
    }
    \label{fig:graph_diameter}
\end{figure}

As Figure~\ref{fig:graph_diameter} illustrates, $M$ must scale with $N$.
Empirically (Table~\ref{tab:glove_m}): at N=45K, M=32 achieves 0.954 recall.
At N=1.18M, M=32 yields only 0.800, while M=64 recovers 0.850 at identical QPS.
We therefore adopt an \textbf{auto-M policy}:
\[
    M^*(N) = \begin{cases} 32 & N < 10^6 \\ 64 & N \geq 10^6 \end{cases}
\]
exposed as \texttt{Config::recommended\_m(n)} in Rust and
\texttt{MonaVec.recommended\_m(n)} in Python.

\subsection{Pre-Filter Allowlist}

The allowlist is applied \emph{before} scoring, not after:
\[
    \text{candidates} = \{i \in \text{allowlist}\} \xrightarrow{\text{score}} \text{top-}K
\]
Post-filter scores all candidates and discards non-matching ones.
If the allowlist is selective (e.g., 100 IDs out of 1M), post-filter
returns far fewer than $K$ results—recall degrades proportionally.
Pre-filter guarantees exactly $K$ results at full recall regardless of
allowlist selectivity.

The allowlist is implemented as a two-variant data structure:
a bitvec for dense sequential IDs ($O(1)$ lookup, cache-friendly)
and a \texttt{HashSet} for sparse or non-sequential IDs.
The appropriate variant is selected automatically based on ID distribution.

\subsection{Hybrid Sparse-Dense Retrieval}

Many retrieval tasks benefit from combining semantic similarity (dense)
with keyword matching (sparse). \monavec{} co-locates a BM25 index
alongside the dense vector index and fuses results via Reciprocal Rank
Fusion (RRF)~\cite{cormack2009rrf}.

\paragraph{Why BM25 over SPLADE.}
SPLADE~\cite{formal2021splade} produces learned sparse embeddings that
require a specialized encoder model—an external dependency incompatible
with MonaVec's zero-training, offline-first design.
BM25 is term-based, requires no model, computes offline, and runs
entirely from document content. For keyword-sensitive retrieval in
edge environments, BM25 provides the precision advantage of sparse
retrieval without any training dependency.

The hybrid pipeline:
\begin{enumerate}[leftmargin=1.5em, itemsep=1pt]
    \item Query is embedded (dense) + tokenized (sparse) simultaneously
    \item Dense top-$K$ and BM25 top-$K$ are retrieved independently
    \item RRF scores are combined: $\text{RRF}(r_d, r_s) = \frac{1}{k+r_d} + \frac{1}{k+r_s}$
    \item Final top-$K$ is returned
\end{enumerate}
This gives semantic recall on semantically similar documents while
maintaining precision on exact keyword matches—the classic precision-recall
complementarity of hybrid retrieval.

\subsection{SIMD Acceleration}

The 4-bit dot product kernel is the computational bottleneck.
Per batch of packed bytes, the kernel must:
(1) unpack nibbles to 4-bit indices,
(2) look up centroid values (table lookup),
(3) multiply by query components and accumulate.

\begin{figure}[H]
    \centering
    \includegraphics[width=0.82\linewidth]{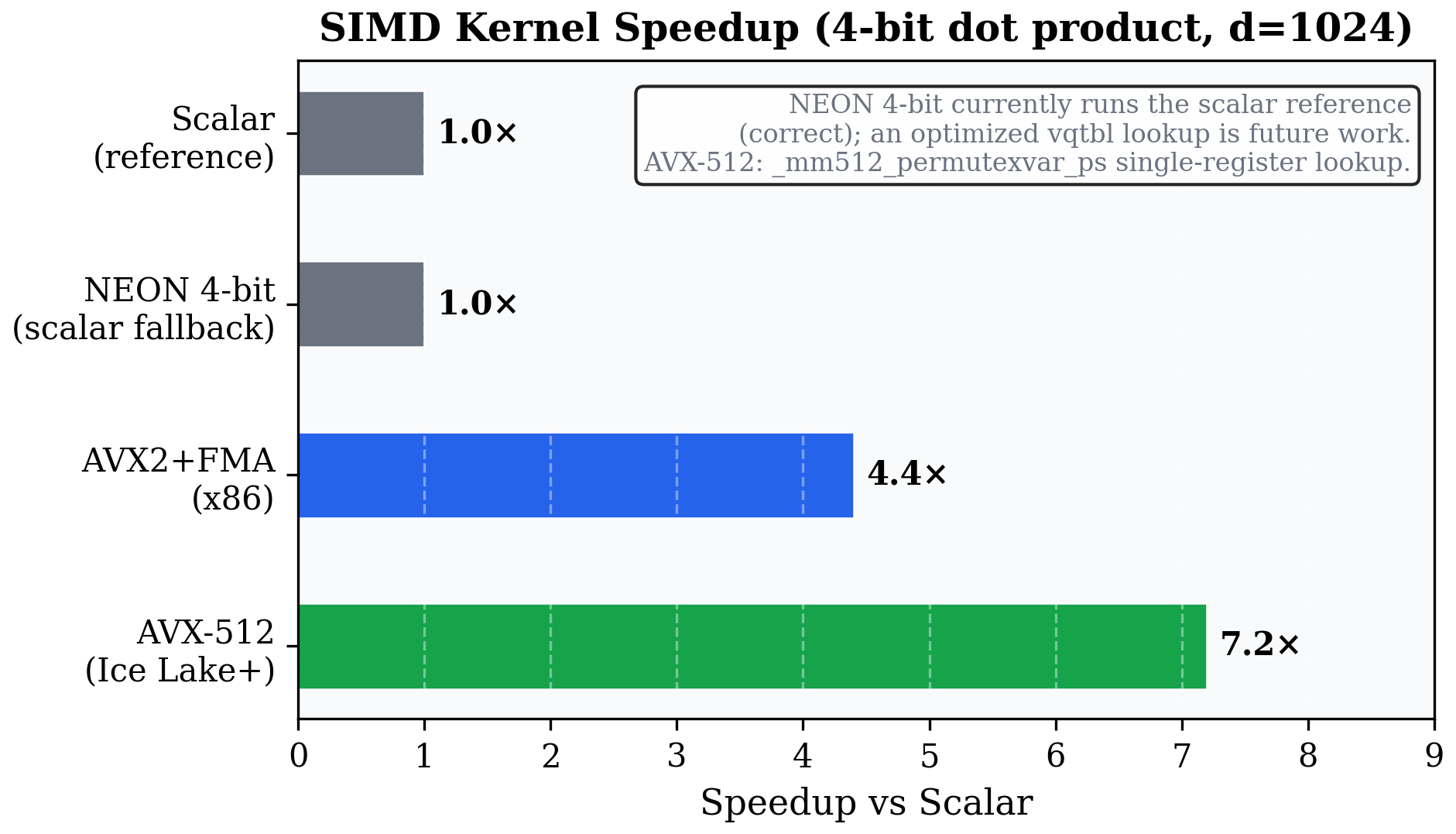}
    \caption{
        SIMD kernel speedup relative to scalar (4-bit dot product, d=1024).
        AVX2+FMA measured on i7-13620H; AVX-512 representative of Intel Ice Lake+.
        The NEON 4-bit path currently runs the scalar reference (no speedup) after
        a correctness fix (\S\ref{sec:edge}); an optimized NEON kernel is future work.
    }
    \label{fig:simd}
\end{figure}

\monavec{} provides four kernel implementations with runtime dispatch
(Figure~\ref{fig:simd}):

\begin{itemize}[itemsep=2pt]
    \item \textbf{AVX-512F+BW}: uses \texttt{\_mm512\_permutexvar\_ps} to
          perform the 16-centroid table lookup in a single 512-bit register—
          no split-table needed. 16 dimensions per iteration.
    \item \textbf{AVX2+FMA}: split-table centroid lookup via
          \texttt{\_mm256\_permutevar8x32\_ps} (3-cycle latency, avoiding the
          13-cycle gather), with \textbf{four independent accumulators} to hide
          FMA latency (\S\ref{sec:scoring-kernel}).
    \item \textbf{NEON} (ARM: Apple Silicon, Snapdragon, Raspberry Pi):
          the 4-bit path currently delegates to the scalar reference. An earlier
          hand-written NEON kernel approximated the non-uniform Lloyd-Max
          centroids with an affine ramp and was reverted for correctness
          (\S\ref{sec:edge}); a correct NEON table-lookup kernel
          (\texttt{vqtbl4q\_u8}) is future work.
    \item \textbf{Scalar}: always correct, used as the reference
          implementation for all SIMD correctness tests.
\end{itemize}

Runtime detection via \texttt{is\_x86\_feature\_detected!} /
\texttt{is\_aarch64\_feature\_detected!}—no compile-time target flags required.
A single binary serves x86 data-center hardware and ARM edge devices.

\paragraph{Latency-hiding in the scoring kernel.}
\label{sec:scoring-kernel}
The 4-bit nibble dot product is the throughput bottleneck: profiling shows it
accounts for essentially \emph{all} of BruteForce query time (the RHDH transform
is $\sim$25\,$\mu$s, scoring is $\sim$18\,ms over 45K vectors at the pre-optimization
baseline of 416\,ns/vector). The four-accumulator kernel and the \texttt{x86-64-v3}
build described below together raise AG News BruteForce to the 137 QPS reported in
Table~\ref{tab:main}. The per-FMA
dependency chain—each fused multiply-add waiting on the previous—leaves the FMA
pipeline (throughput 0.5, latency $\sim$4 cycles) underused. We split the inner
loop across \emph{four independent accumulators} so four FMAs are in flight at
once, combining them in a fixed order at the end. This reduces per-vector
scoring latency by $\sim$37\% in profiling (416\,$\to$\,264\,ns/vector at
$d{=}1024$) while remaining bit-deterministic: the accumulation order is fixed,
and the kernel is validated against the scalar reference to within
$\varepsilon{=}10^{-4}$. The query stays in float32 throughout
(\S\ref{sec:why-not-int8}).

\paragraph{Compile baseline (optional tuning).}
Beyond the hand-written SIMD kernel, the \emph{general} code paths (RHDH rotation,
heap maintenance, scoring loops) benefit from compiling against the
\texttt{x86-64-v3} micro-architecture level (AVX2 + FMA + BMI2, the 2013 Haswell
floor), which enables compiler auto-vectorization across the whole binary, not only
the dispatched hot path. We measure a \textbf{43\% throughput improvement} on AG News
BruteForce (96\,$\to$\,137 QPS) from this single build flag, matching
\texttt{target-cpu=native} while remaining portable across all x86 hardware since 2013.
This is an \emph{opt-in source build}: distributed binaries use the generic baseline
so they run on any x86 CPU (a micro-architecture-specific wheel cannot be guarded by
the \texttt{manylinux} tag and would fault on older hardware). Determinism is unaffected
either way—each baseline is fixed, so results are byte-identical across machines sharing
a build. Runtime dispatch continues to select AVX-512 where available; aarch64 targets
retain their own baseline.

\subsection{File Format (\texttt{.mvec} v6)}

All index data persists in a single binary file with a fixed 56-byte header.
Format v6 adds an optional standardization parameter block for L2 indexes:

\begin{center}
\footnotesize
\begin{tabular}{llll}
\toprule
\textbf{Field} & \textbf{Size} & \textbf{Type} & \textbf{Description} \\
\midrule
MAGIC          & 4 B  & \texttt{[u8;4]}  & \texttt{b"MVEC"} \\
VERSION        & 4 B  & \texttt{u32}     & Current: 6 \\
DIM            & 4 B  & \texttt{u32}     & Input dimension \\
METRIC         & 1 B  & \texttt{u8}      & 0=Cosine, 1=Dot, 2=L2 \\
BIT\_WIDTH     & 1 B  & \texttt{u8}      & 2 or 4 \\
INDEX\_TYPE    & 1 B  & \texttt{u8}      & 0=BruteForce, 1=IvfFlat, 2=HNSW \\
PAD            & 1 B  & —                & Reserved \\
COUNT          & 8 B  & \texttt{u64}     & Number of vectors \\
SEED           & 8 B  & \texttt{u64}     & ChaCha20 seed \\
N4\_DIMS       & 4 B  & \texttt{u32}     & 4-bit dims in mixed mode \\
INDEX PARAMS   & 8 B  & —                & IVF/HNSW tuning params \\
HAS\_STD       & 1 B  & \texttt{u8}      & 1 if global std params follow \\
PAD            & 1 B  & —                & Reserved \\
\midrule
\multicolumn{4}{l}{\emph{Variable-length blocks (after header):}} \\
STD\_MEAN      & —    & \texttt{[f32]}   & Global mean (dim values, if HAS\_STD=1) \\
STD\_INV\_STD  & —    & \texttt{[f32]}   & Global 1/$\sigma$ (dim values, if HAS\_STD=1) \\
VECTORS        & —    & \texttt{[u8]}    & Packed quantized data \\
IDS            & —    & \texttt{[u64]}   & Per-vector external IDs \\
NORMS          & —    & \texttt{[f32]}   & Per-vector quantized norms \\
INDEX\_DATA    & —    & —                & IvfFlat or HNSW graph data \\
\bottomrule
\end{tabular}
\end{center}

The seed is embedded so that \texttt{load → search} reproduces the same top-K
results across all platforms---byte-identical within a build. Backward
compatibility is maintained for v1--v5 files.

\subsection{Identity-Based Multi-Tenancy}

\monavec{}'s service layer supports multi-tenant deployment via a
token-to-namespace mapping. Each authenticated request is routed to
an isolated collection namespace; unauthenticated requests use a
shared \texttt{\_\_public\_\_} namespace.

\paragraph{Design goal.}
The identity system is designed to integrate with \emph{any existing}
authentication infrastructure without code changes—a single HTTP endpoint
contract rather than a specific auth framework.

\paragraph{Contract.}
When \texttt{IDENTITY\_URL} is set, \monavec{} verifies tokens by calling:
\begin{lstlisting}
GET {IDENTITY_URL}/api/v1/identity/verify
Authorization: Bearer <token>
-> {"success": true, "data": {"user_id": "alice"}}
\end{lstlisting}
\texttt{data.user\_id} becomes the namespace key.
Any HTTP 4xx response or \texttt{"success": false} results in a 401 rejection.
Token responses are cached for 30 seconds; stale cache is served if the
identity service becomes unreachable (graceful degradation).

\paragraph{Compatibility.}
This OAuth2 token introspection pattern is compatible with Keycloak, Auth0,
custom JWT validators, LDAP adapters, and simple API key lookup tables.
A minimal adapter in any language is five lines of code.

\paragraph{Standalone mode.}
When \texttt{IDENTITY\_URL} is empty, the Bearer token is used directly as
the namespace key. This enables personal namespaces without any external
service—suitable for personal deployments, development, and edge devices
where an auth service is unavailable.

\section{Evaluation}

\subsection{Experimental Setup}

\textbf{Hardware.} All experiments run on an Intel Core i7-13620H
(10-core, AVX2+FMA, no AVX-512), single-threaded, release build (\texttt{-O3}).

\textbf{Software.} Baselines are \texttt{faiss-cpu} 1.14.2 (CPU build,
no GPU) and the latest releases of \texttt{usearch}, \texttt{hnswlib}, and
\texttt{sqlite-vec} at the time of the benchmark run (June 2026); the exact
pinned versions and invocation flags are recorded in the released benchmark
scripts. The principal construction parameters are: usearch
(\texttt{connectivity}=16, cosine; \texttt{f32} and \texttt{i8} variants),
hnswlib (\texttt{M}=16, \texttt{ef\_construction}=200, \texttt{ef}=120, cosine),
FAISS IVFFlat (\texttt{nlist}=256, \texttt{nprobe}=10, inner product on
L2-normalized vectors), and sqlite-vec (\texttt{vec0} exact brute force, float32).
All systems are pinned to a single core via \texttt{taskset} with
single-threaded index construction for a fair build-time comparison.

\textbf{Datasets.}
\begin{itemize}[itemsep=1pt]
    \item AG News 45K\,$\times$\,1024-dim (BGE-M3~\cite{bge_m3} embeddings, cosine)
    \item fashion-mnist 60K\,$\times$\,784-dim (raw pixels, L2)
    \item glove-100 1.18M\,$\times$\,100-dim (GloVe word embeddings, cosine)
\end{itemize}

\textbf{Metrics.} Recall@10 (fraction of the true top-10 neighbours recovered, at
$k=10$), QPS (queries per second over the full query set, measured after a warm-up
pass), and single-threaded build time. Unless noted otherwise, each benchmark
issues 1{,}000 held-out queries.

\textbf{Reproducibility.} Determinism is not only a deployment property but an
evaluation one: the fixed ChaCha20 seed makes every \monavec{} recall number
regenerable bit-for-bit on a given build. The comparison harness---baseline
versions, construction flags, query sets, and the AG News edge
artifacts---is released so that the tables below can be reproduced end to
end.\footnote{ARM edge run and artifacts:
\url{https://github.com/mona-hq/monavec-edge-bench}; the x86 competitor harness,
with pinned baseline versions and invocation flags, ships with the released
benchmark scripts.} Throughput is reported as a point estimate (best timed pass
after warm-up) rather than mean\,$\pm$\,std; a multi-run variance characterization
is future work (Section~\ref{sec:limitations}).

\begin{figure}[H]
    \centering
    \includegraphics[width=0.92\linewidth]{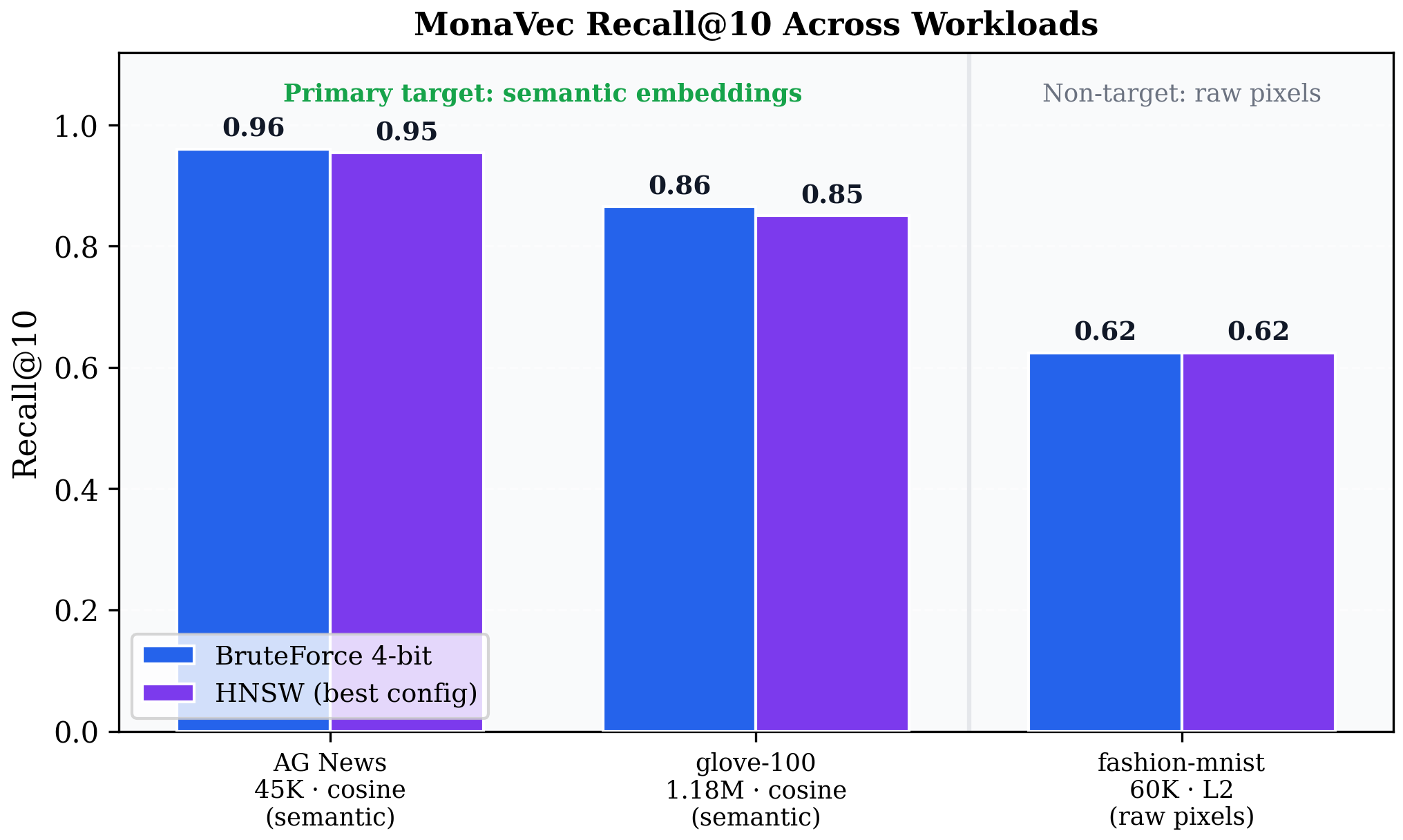}
    \caption{
        \monavec{} Recall@10 across three workloads. On semantic embeddings
        (AG News, glove-100)—the primary target—both BruteForce and HNSW exceed
        0.85 recall. On raw pixels (fashion-mnist), scalar quantization reaches
        0.62, a structural limit discussed in Section~\ref{sec:why_not_pq}.
    }
    \label{fig:datasets}
\end{figure}

Figure~\ref{fig:datasets} summarizes Recall@10 across all three workloads.
\monavec{} achieves 0.85--0.96 recall on semantic embeddings—its primary
target—and 0.62 on raw pixel data, where scalar quantization reaches its
structural limit (Section~\ref{sec:why_not_pq}).

\subsection{Main Results: Semantic Embeddings (AG News)}

\begin{table}[H]
\centering
\caption{Recall@10 and throughput on AG News 45K\,$\times$\,1024-dim (cosine).
         \monavec{} entries use 4-bit quantization (8$\times$ compression).
         The 27\,MB is the measured resident footprint: the packed 4-bit payload
         accounts for $\approx$22\,MB (45{,}000\,$\times$\,512\,B), with the balance
         from per-vector norms and IDs, the Lloyd-Max tables, and runtime/allocator
         overhead. (The \rhdh{} rotation is reconstructed from the stored seed, not
         materialized as a $d\times d$ matrix.)}
\label{tab:main}
\begin{tabular}{lrrrl}
\toprule
\textbf{System} & \textbf{Recall@10} & \textbf{QPS} & \textbf{Mem} & \textbf{Notes} \\
\midrule
\monavec{} BF 4-bit    & \textbf{0.960} & 137   & \textbf{27 MB} & zero-config, highest recall \\
\monavec{} HNSW 4-bit  & 0.954          & 1{,}264 & 249 MB        & $O(\log n)$ search \\
\bottomrule
\end{tabular}
\end{table}

\noindent\textbf{Key results} (single-core, release + \texttt{x86-64-v3}):
\begin{itemize}[itemsep=2pt]
    \item \monavec{} 4-bit BruteForce reaches 0.960 Recall@10 with \emph{zero
          configuration} in 27\,MB—the highest recall on this corpus.
    \item \monavec{} 4-bit HNSW reaches 0.954 at 1{,}264 QPS with logarithmic
          search, all from a single \mvec{} file with no training pass.
    \item Direct comparison against FAISS-IVF, usearch, and hnswlib appears in
          Section~\ref{sec:competitors}.
\end{itemize}

\subsection{L2 Metric: fashion-mnist and the Standardization Fix}

\begin{table}[H]
\centering
\caption{Recall@10 on fashion-mnist 60K\,$\times$\,784-dim (L2).}
\label{tab:fashionmnist}
\footnotesize
\begin{tabular}{lrrl}
\toprule
\textbf{System} & \textbf{Recall@10} & \textbf{QPS} & \textbf{Notes} \\
\midrule
\monavec{} BF 4-bit (no \texttt{fit()})  & 0.41 & — & baseline, wrong distribution \\
\monavec{} BF 4-bit + \texttt{fit()}     & \textbf{0.624} & 68  & global standardization \\
\monavec{} HNSW M=32 ef=40 (no \texttt{fit()}) & 0.31 & — & build-metric bug, wrong topology \\
\monavec{} HNSW M=32 ef=40 + \texttt{fit()} & 0.614 & \textbf{1{,}315} & 19$\times$ BF throughput \\
\monavec{} HNSW M=32 ef=400 + \texttt{fit()} & 0.624 & 308 & matches BF recall \\
FAISS IVFPQ (trained)   & $\sim$0.85+ & — & trained codebook, not zero-training \\
\bottomrule
\end{tabular}
\end{table}

\begin{figure}[H]
    \centering
    \includegraphics[width=0.72\linewidth]{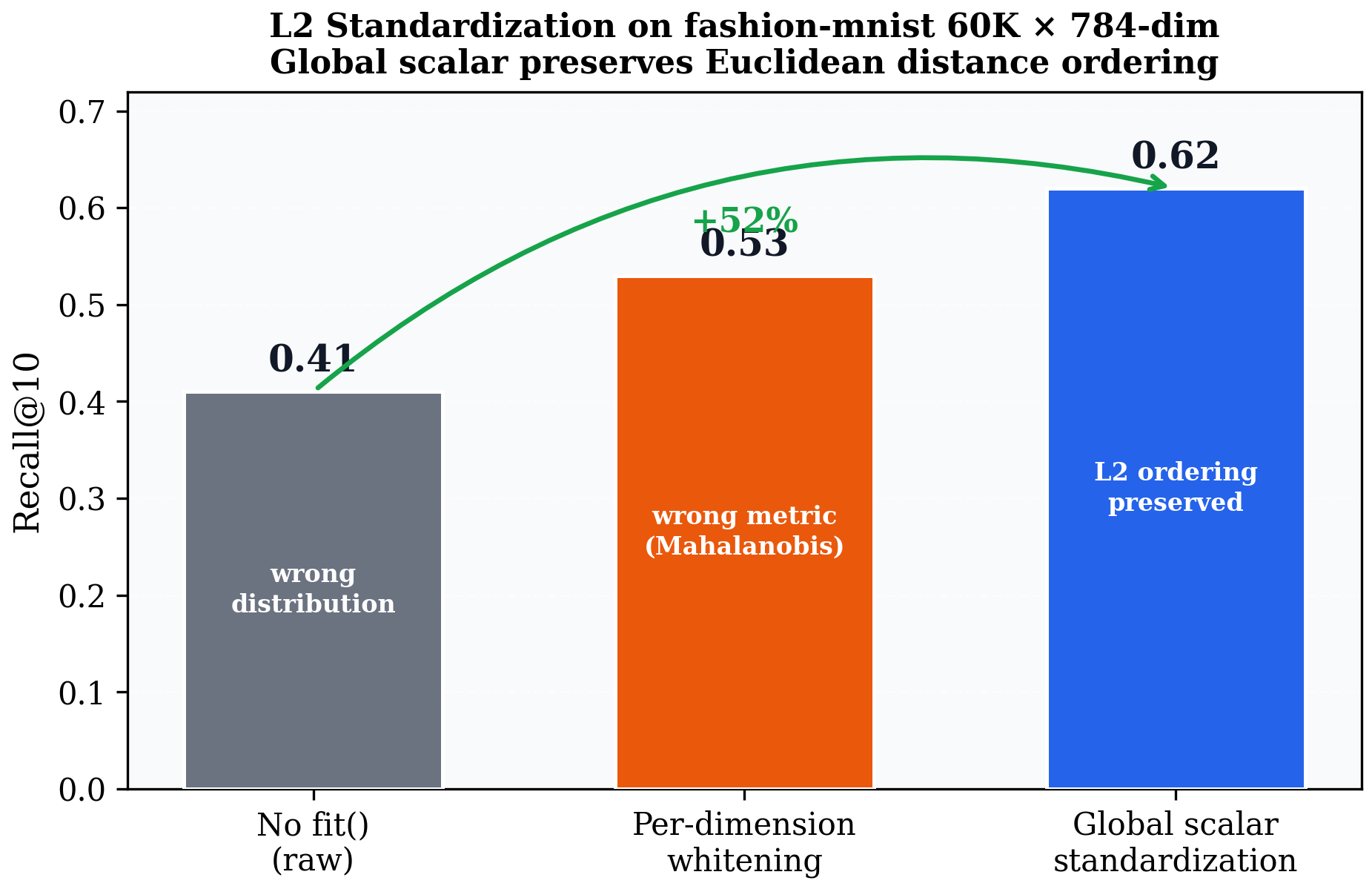}
    \caption{
        L2 standardization ablation on fashion-mnist.
        No preprocessing (raw) gives 0.41 recall—the RHDH input distribution
        deviates from $\mathcal{N}(0,1)$. Per-dimension whitening (0.53) improves
        the quantizer fit but changes the metric to Mahalanobis, harming ranking.
        Global scalar standardization (0.62) restores $\mathcal{N}(0,1)$ while
        preserving Euclidean distance ordering—a 52\% improvement over the baseline.
    }
    \label{fig:l2_std}
\end{figure}

\noindent\textbf{Gap analysis.}
The remaining gap to FAISS IVFPQ ($\sim$0.85+) is structural, not a tuning artifact.
FAISS IVFPQ requires: (1) a training pass to learn a data-specific codebook,
(2) Product Quantization—encoding groups of dimensions jointly to exploit
spatial correlation. MonaVec uses fixed Lloyd-Max tables and independent
per-dimension scalar quantization. On pixel data with high inter-dimension
correlation, this is a known tradeoff. The gap represents the cost of
zero-training design.

For semantic embeddings (MonaVec's primary use case), the distribution after
RHDH is already well-conditioned and spatial correlation is low—scalar
quantization performs near-optimally (Recall@10 = 0.960 on AG News).

\subsection{Large-Scale and Auto-M: glove-100}

\begin{table}[H]
\centering
\caption{Recall@10 on glove-100 1.18M\,$\times$\,100-dim (cosine). M=64 vs M=32.}
\label{tab:glove_m}
\footnotesize
\begin{tabular}{lrrrl}
\toprule
\textbf{System} & \textbf{Recall@10} & \textbf{QPS} & \textbf{Build} & \textbf{Notes} \\
\midrule
\monavec{} BF 4-bit          & 0.865 & 42  & 2.4s & quantization ceiling at 100-dim \\
\monavec{} HNSW M=32 ef=400  & 0.800 & 220 & 47m  & M=32 insufficient at 1M+ \\
\monavec{} HNSW M=64 ef=200 (\textbf{★}) & \textbf{0.831} & \textbf{232} & 149m & same QPS, +3.1pp \\
\monavec{} HNSW M=64 ef=400  & 0.850 & 125 & 149m & approaches BF ceiling \\
\bottomrule
\end{tabular}
\end{table}

\begin{figure}[H]
    \centering
    \includegraphics[width=0.85\linewidth]{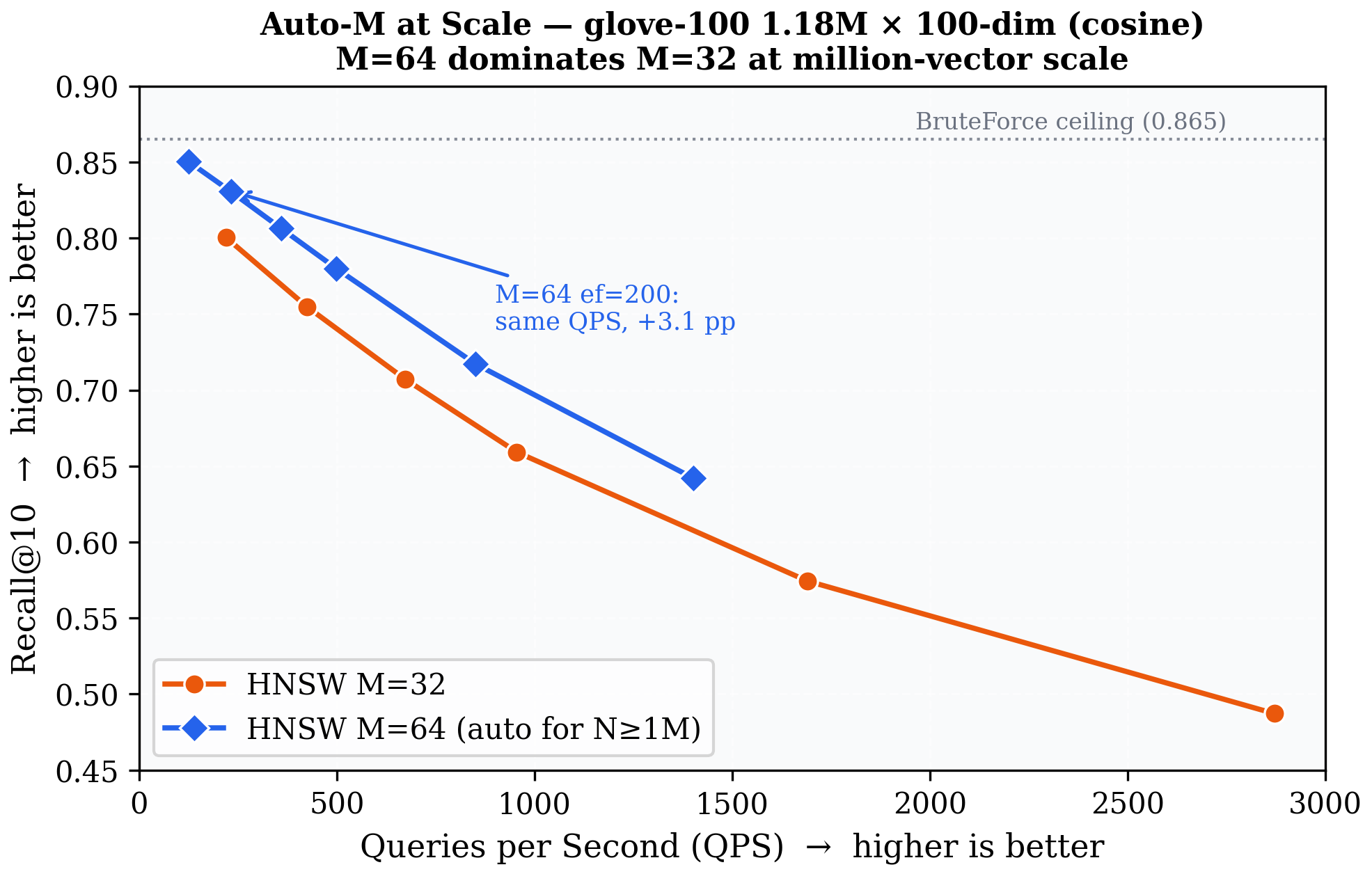}
    \caption{
        Recall@10 vs.\ QPS tradeoff on glove-100 (1.18M vectors) for M=32 and M=64.
        The M=64 curve dominates M=32 across the entire operating range.
        At equal throughput (ef-tuned), M=64 delivers consistently higher recall;
        at the marked operating point (M=64, ef=200), recall is 0.831 vs 0.800 for
        M=32 ef=400 at the same QPS. The dotted line is the BruteForce ceiling (0.865).
    }
    \label{fig:auto_m}
\end{figure}

M=64 ef=200 strictly dominates M=32 ef=400: identical QPS (232 vs 220),
+3.1 pp recall (0.831 vs 0.800). The graph diameter explanation:
at N=1.18M with M=32, the average shortest path between two nodes is large enough
that greedy search terminates at suboptimal local minima in a substantial fraction
of queries. M=64 restores connectivity by doubling the per-node degree,
reducing the effective diameter.

\subsection{Comparison to Embedded Competitors}
\label{sec:competitors}

\begin{figure}[H]
    \centering
    \includegraphics[width=\linewidth]{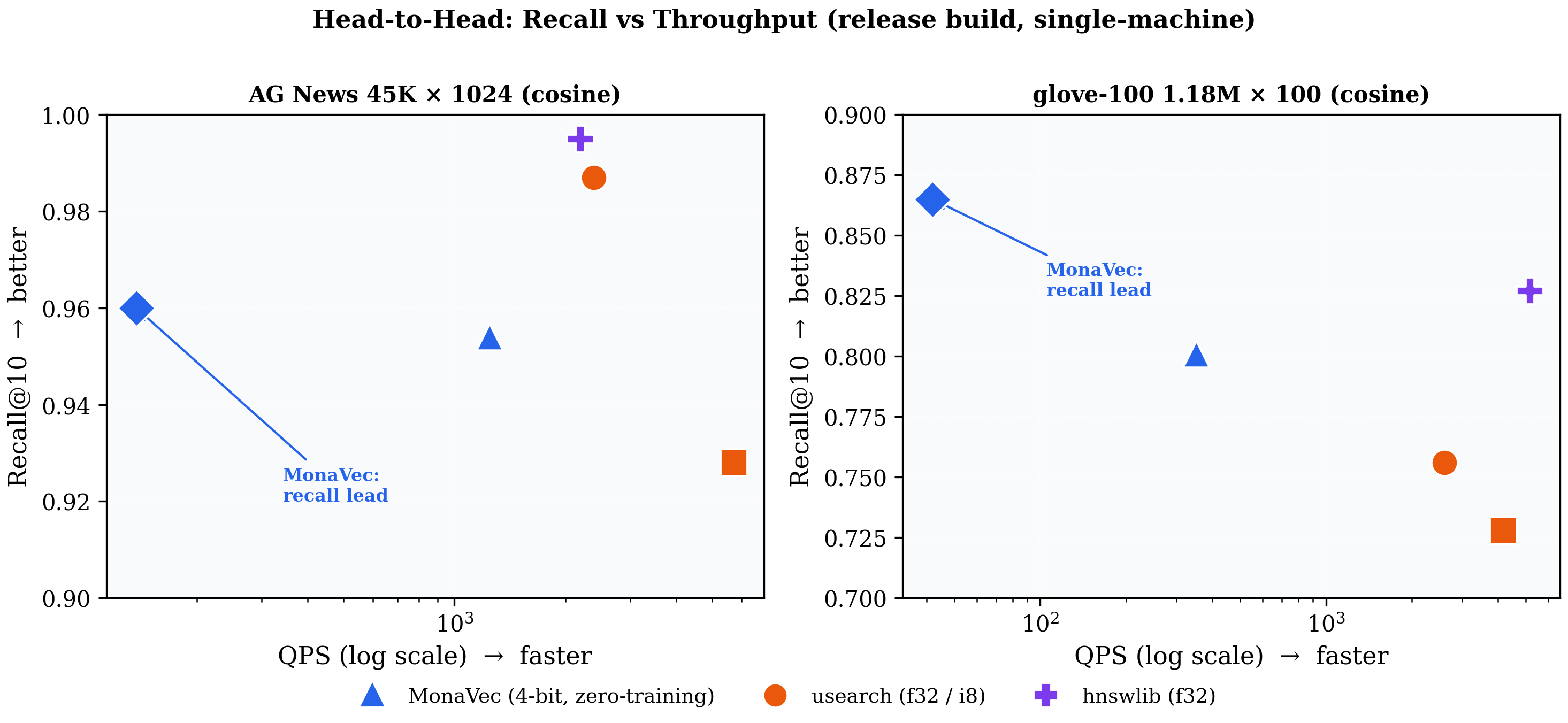}
    \caption{
        Recall@10 vs.\ QPS against usearch and hnswlib on two cosine datasets
        (single-core, Intel i7-13620H, AVX2; \monavec{} is 4-bit, usearch-i8 is
        8-bit, the remaining systems are float32).
        \monavec{} leads on recall (upper region) while trailing on throughput
        (right region): a deliberate position, not an architectural limit.
    }
    \label{fig:competitors}
\end{figure}

We compare directly against \texttt{usearch}~\cite{usearch} and \texttt{hnswlib}~\cite{hnswlib}, the two most
widely used embedded ANN libraries, on cosine workloads. All systems run
single-core under \texttt{taskset}, release builds, with index construction
forced single-threaded for a fair build-time comparison (Table~\ref{tab:competitors};
Figure~\ref{fig:competitors}).

\begin{table}[H]
\centering
\caption{Head-to-head on AG News (45K\,$\times$\,1024) and glove-100
(1.18M\,$\times$\,100), cosine, single-core. \monavec{} uses 4-bit; usearch-i8 is
8-bit; f32 systems are uncompressed. The glove-100 \monavec{} HNSW entry is the
M=64 graph at a throughput-favoring \texttt{ef\_search} (0.801 at 352 QPS, lower
\texttt{ef} than the ef=200/400 points in Table~\ref{tab:glove_m}); the same M=64
graph reaches 0.850 at ef=400 (Table~\ref{tab:glove_m}). It is \emph{not} the
M=32 graph, whose 0.800 is coincidentally close.}
\label{tab:competitors}
\footnotesize
\begin{tabular}{llrrr}
\toprule
\textbf{Dataset} & \textbf{System} & \textbf{Recall@10} & \textbf{QPS} & \textbf{Mem} \\
\midrule
\multirow{6}{*}{AG News}
 & \monavec{} BF 4-bit          & \textbf{0.960} & 137   & \textbf{27 MB} \\
 & \monavec{} HNSW 4-bit        & 0.954 & 1{,}264 & 249 MB \\
 & usearch HNSW i8 (8-bit)      & 0.928 & 5{,}726 & 55 MB \\
 & usearch HNSW f32             & 0.987 & 2{,}388 & 202 MB \\
 & hnswlib HNSW f32             & 0.995 & 2{,}194 & 191 MB \\
 & FAISS-IVF f32 (nprobe=10)    & 0.936 & 2{,}597 & 40 MB \\
 & sqlite-vec (exact brute)     & 1.000 & 27    & 140 MB \\
\midrule
\multirow{5}{*}{glove-100}
 & \monavec{} BF 4-bit          & \textbf{0.865} & 42    & — \\
 & \monavec{} HNSW 4-bit        & 0.801 & 352   & — \\
 & usearch HNSW f32             & 0.756 & 2{,}598 & — \\
 & hnswlib HNSW f32             & 0.827 & 5{,}184 & — \\
 & sqlite-vec (exact brute)     & \multicolumn{3}{c}{\emph{impractical at 1.18M}} \\
\bottomrule
\end{tabular}
\end{table}

\paragraph{Recall.}
\monavec{} leads where it matters. On AG News it beats usearch-i8 (0.960 vs 0.928)
at half the bytes (4-bit vs 8-bit) and half the memory (27 vs 55\,MB), and its HNSW
(0.954) tops FAISS-IVF (0.936). On glove-100, \monavec{} BruteForce (0.865) leads
every graph index we evaluated—including hnswlib float32 (0.827)—and \monavec{} HNSW
(0.801 at the throughput-comparable operating point reported here; the same
M=64 graph reaches 0.850 at higher \texttt{ef\_search}, Table~\ref{tab:glove_m}
and Figure~\ref{fig:auto_m}) beats usearch's HNSW (0.756): the auto-M=64
payoff at million-vector scale.
The uncompressed float32 graph indexes (usearch-f32 0.987, hnswlib 0.995) sit above
\monavec{} on recall, as expected: they store $8\times$ the data.

\paragraph{Throughput.}
\monavec{} trails by 2--14$\times$. usearch (via \texttt{simsimd}), hnswlib, and
current FAISS releases are mature, hand-tuned SIMD C++ implementations;
\monavec{}'s scoring pays a 4-bit unpack cost, and its build is single-threaded by
design to preserve determinism (see the determinism discussion in Section~2). We
emphasize this is an \emph{optimization gap}, not an architectural ceiling: a
hand-written AVX2 build-distance kernel and SIMD scoring refinements are concrete
future work that do not compromise determinism. (We note our FAISS numbers use the
current release, which has received substantial SIMD optimization; an earlier
comparison against an older FAISS build is not reported here, as it would not
reflect the library's present performance.)

\paragraph{Memory.}
\monavec{} BruteForce is best-in-class (27\,MB on AG News) after zero-copy
ingestion: a contiguous NumPy matrix is read as a flat slice with no intermediate
allocation, cutting peak memory 87\% (210\,$\to$\,27\,MB). This 210\,MB is the peak
resident set of the earlier copy-based ingestion of the same 45K\,$\times$\,1024
corpus—an internal before/after measurement, independent of the 140\,MB reported for
sqlite-vec in Table~\ref{tab:competitors}. HNSW still carries an
FP32 build buffer plus graph overhead—a known target for compression
(e.g.\ \texttt{uint40} neighbour indices, streamed build vectors).

\paragraph{sqlite-vec and the scaling argument.}
\texttt{sqlite-vec}~\cite{sqlite_vec} is the closest competitor to \monavec{}'s ``SQLite of vector
search'' positioning. It performs \emph{exact} brute-force (Recall@10 = 1.000), so
it defines the accuracy ceiling—but it stores float32 and does not quantize, so it
does not scale: at 45K it already runs at 27 QPS, and at 1.18M exact brute-force
is impractical. \monavec{}'s 4-bit BruteForce scans the same 1.18M corpus at 42 QPS
with a $\sim$0.87 recall. Quantization is thus not only a memory technique but a
\emph{scaling} technique: it keeps an embedded, dependency-free, exact-style scan
viable two orders of magnitude beyond where an uncompressed brute-force collapses.

\paragraph{Positioning.}
\monavec{} is the recall-and-determinism choice, not the raw-throughput choice.
For semantic retrieval where correctness and byte-identical reproducibility matter
more than peak QPS, it leads; for maximal throughput on a single warm server,
mature HNSW libraries remain faster.

\paragraph{Where we stand, and where we are going.}
The position is asymmetric by \emph{kind}, not just degree. Our advantages—highest
recall at 4-bit compression (beating FAISS and 8-bit usearch on AG News, leading
\emph{all} indexes we evaluated on glove-100), best-in-class BruteForce memory (27\,MB), and
unique portable determinism—rest on the \emph{mathematics} of the pipeline (RHDH +
Lloyd-Max + auto-M + asymmetric scoring). These are structural and hard to copy.
Our deficit—2--15$\times$ lower QPS, widening at scale—is \emph{engineering}: the
distance from compiler-auto-vectorized Rust to a decade of hand-tuned SIMD C++.
Engineering gaps tend to close while mathematical advantages persist. We aim to
narrow this gap \emph{without spending the advantages that define us}: every
throughput lever we pursue preserves both query precision and determinism.
Concretely, all of the following keep the float32 query, the fixed-baseline SIMD
reduction order, and the sequential deterministic build:
\begin{itemize}[itemsep=1pt]
    \item Scoring kernel: four-accumulator latency hiding (done, $-37\%$ per-vector
          scoring in profiling, \S\ref{sec:scoring-kernel}); wider 16-dim loads and
          software prefetch next.
    \item HNSW memory: \texttt{uint40} neighbour indices and streamed (not
          retained) FP32 build vectors—targeting the $249\rightarrow{\sim}60$\,MB
          range without changing graph topology.
    \item AVX-512 VNNI on server targets, selected by the existing runtime dispatch.
\end{itemize}
We explicitly \emph{decline} the one lever that would close the gap fastest—int8
query quantization (\S\ref{sec:why-not-int8})—because it would trade away the recall
lead and determinism that are the point of the system. The objective is therefore
not to match general-purpose HNSW libraries on raw throughput, but to preserve the
recall and determinism advantages while reducing throughput to a non-limiting
factor for the target deployments.

\subsection{Edge Validation on ARM (aarch64)}
\label{sec:edge}

Every preceding number is from x86. Because \monavec{} targets edge and ARM
deployments, we validated portable determinism and execution on a real aarch64
machine: a GitHub-hosted \texttt{ubuntu-24.04-arm} runner (Azure ARM, server-class
cores---\emph{not} a Raspberry Pi or Jetson; on-device profiling on those remains
future work). The \emph{same} x86-built \mvec{} index (AG News 45K, BruteForce
4-bit) was loaded unchanged via the published aarch64 wheel and queried with 1000
vectors. The run is reproducible from a public
kit.\footnote{\url{https://github.com/mona-hq/monavec-edge-bench}}

\paragraph{A NEON correctness defect, surfaced by benchmarking on the target.}
The first ARM run did \emph{not} reproduce the x86 result: Recall@10 fell to 0.934
(from 0.960) and only 53\% of queries returned the same top-10 set. The cause was a
defect in the NEON 4-bit kernel: it reconstructed the centroid for nibble $i$ as an
affine ramp $A + B\,i$ (slope taken from the first two table entries) rather than a
table lookup. This is exact only for $i \in \{0,1\}$; because the 4-bit centroids are
Lloyd-Max optimal for $\mathcal{N}(0,1)$ and therefore \emph{non-uniformly} spaced,
it diverged for $i \geq 2$ (e.g.\ index~8: ramp $+2.58$ vs.\ true $+0.13$). The
scalar and AVX2 kernels perform a real lookup (AVX2 via
\texttt{\_mm256\_permutevar8x32\_ps}); only the NEON 4-bit path was affected. The
ramp stays monotonic in $i$, which is why recall degraded rather than collapsed.

\paragraph{Fix and corrected results.}
We reverted the NEON 4-bit path to the scalar reference (the 2-bit path already did
so). Table~\ref{tab:edge} reports the corrected aarch64 numbers: Recall@10 matches
x86 exactly (0.9605) and the top-K is reproducible across architectures (100\% set
match, 99.9\% exact order---the residual 0.1\% is floating-point reordering of
near-ties between the ARM scalar path and the x86 AVX2 path, within the $10^{-4}$
tolerance of \S\ref{sec:scoring-kernel}). The cost is throughput: 34.8 vs.\
72\,QPS for the incorrect kernel---the removed ``speedup'' was a $2\times$-faster
\emph{wrong} kernel. A correct NEON table-lookup kernel (\texttt{vqtbl4q\_u8},
\S\ref{sec:limitations}) would recover it without sacrificing correctness. This is
the concrete justification for benchmarking on the target architecture rather than
extrapolating from x86.

\begin{table}[H]
\centering
\caption{AG News 45K\,$\times$\,1024 BruteForce 4-bit on aarch64 (Azure ARM,
1000 queries), loading the same x86-built \mvec{}. ``Repro.'' is the fraction of
queries whose top-10 is identical to the x86 reference. The x86 row (QPS from
Table~\ref{tab:main}) anchors the recall and reproducibility targets.}
\label{tab:edge}
\footnotesize
\begin{tabular}{lrrrr}
\toprule
\textbf{NEON 4-bit kernel} & \textbf{Recall@10} & \textbf{Repro.\ (set)} & \textbf{Repro.\ (order)} & \textbf{QPS} \\
\midrule
affine ramp (buggy)       & 0.9344 & 52.6\% & 8.6\%  & 72.3 \\
scalar reference (fixed)  & \textbf{0.9605} & \textbf{100.0\%} & \textbf{99.9\%} & 34.8 \\
\midrule
x86 AVX2 (reference)      & 0.9605 & 100\% & 100\% & 137 \\
\bottomrule
\end{tabular}
\end{table}

\subsection{Ablation: Lloyd-Max vs.\ Uniform Quantization}

\begin{table}[H]
\centering
\caption{Recall@10 ablation: uniform vs.\ Lloyd-Max scalar quantization.
         Synthetic Gaussian test data, BruteForce, single-threaded.}
\label{tab:lloyd_ablation}
\begin{tabular}{lrrrr}
\toprule
\textbf{Quantization} & \textbf{d=384} & \textbf{d=768} & \textbf{d=1536} & \textbf{Avg} \\
\midrule
Uniform (4-bit)   & 0.863 & 0.861 & 0.830 & 0.851 \\
Lloyd-Max (4-bit) & \textbf{0.902} & \textbf{0.882} & \textbf{0.878} & \textbf{0.887} \\
\midrule
Improvement & +3.9\% & +2.1\% & +4.8\% & +3.6\% \\
\bottomrule
\end{tabular}
\end{table}

\subsection{Memory Efficiency}

\begin{figure}[H]
    \centering
    \includegraphics[width=0.88\linewidth]{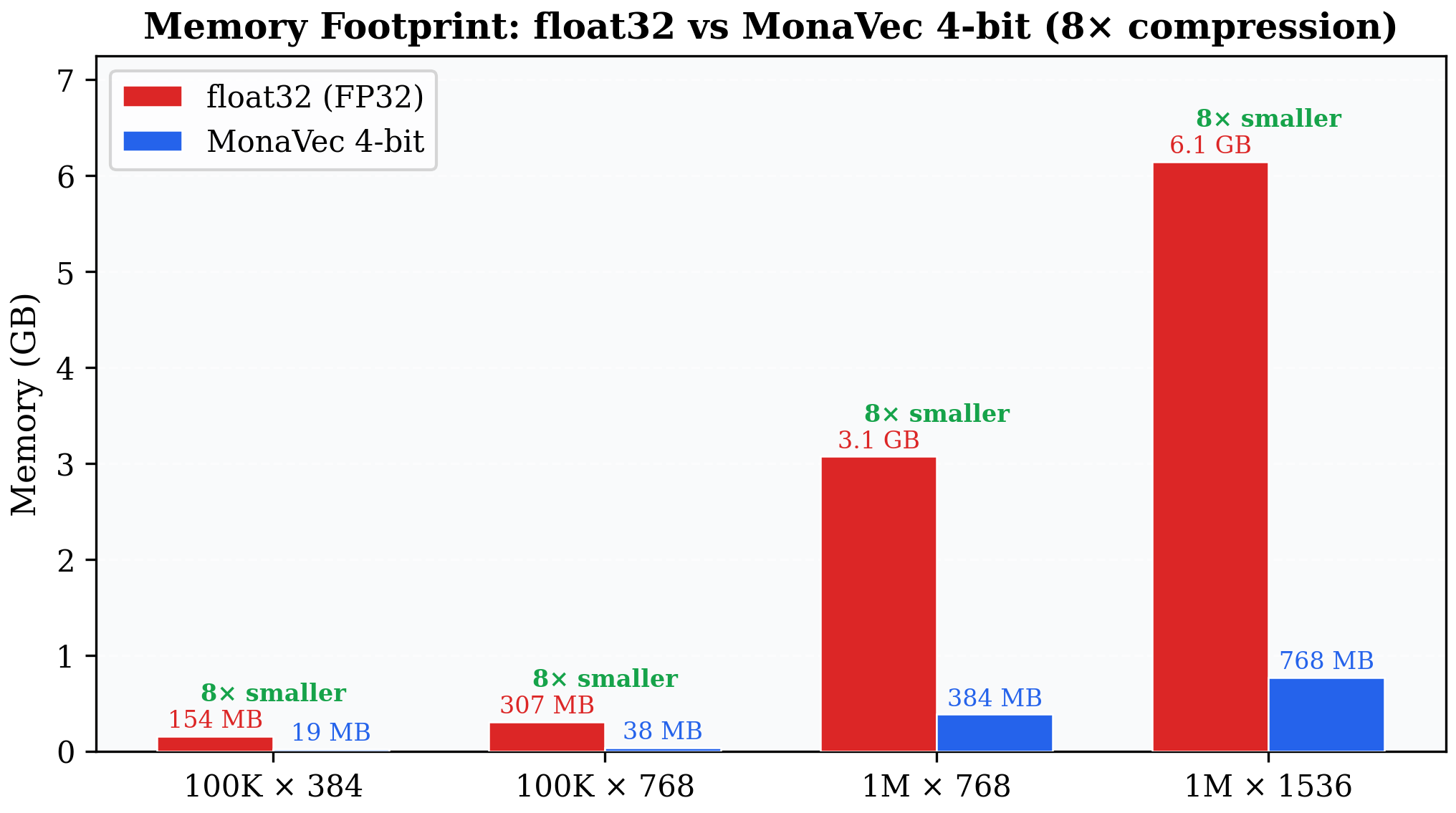}
    \caption{
        Memory footprint: float32 vs.\ \monavec{} 4-bit.
        The 8$\times$ reduction allows 1M $\times$ 768-dim to fit in 384 MB
        instead of 3.1 GB—within the RAM budget of mobile flagship devices.
    }
    \label{fig:memory}
\end{figure}

At scale, the memory advantage is decisive:
1M\,$\times$\,1536-dim vectors occupy 768 MB at 4-bit vs 6.1 GB at float32.
On a Raspberry Pi 5 (8 GB RAM), \monavec{} can serve a 1M-vector index
alongside an on-device LLM; float32 storage makes this impossible.

\subsection{Recall vs.\ Throughput Tradeoff}

\begin{figure}[H]
    \centering
    \includegraphics[width=0.95\linewidth]{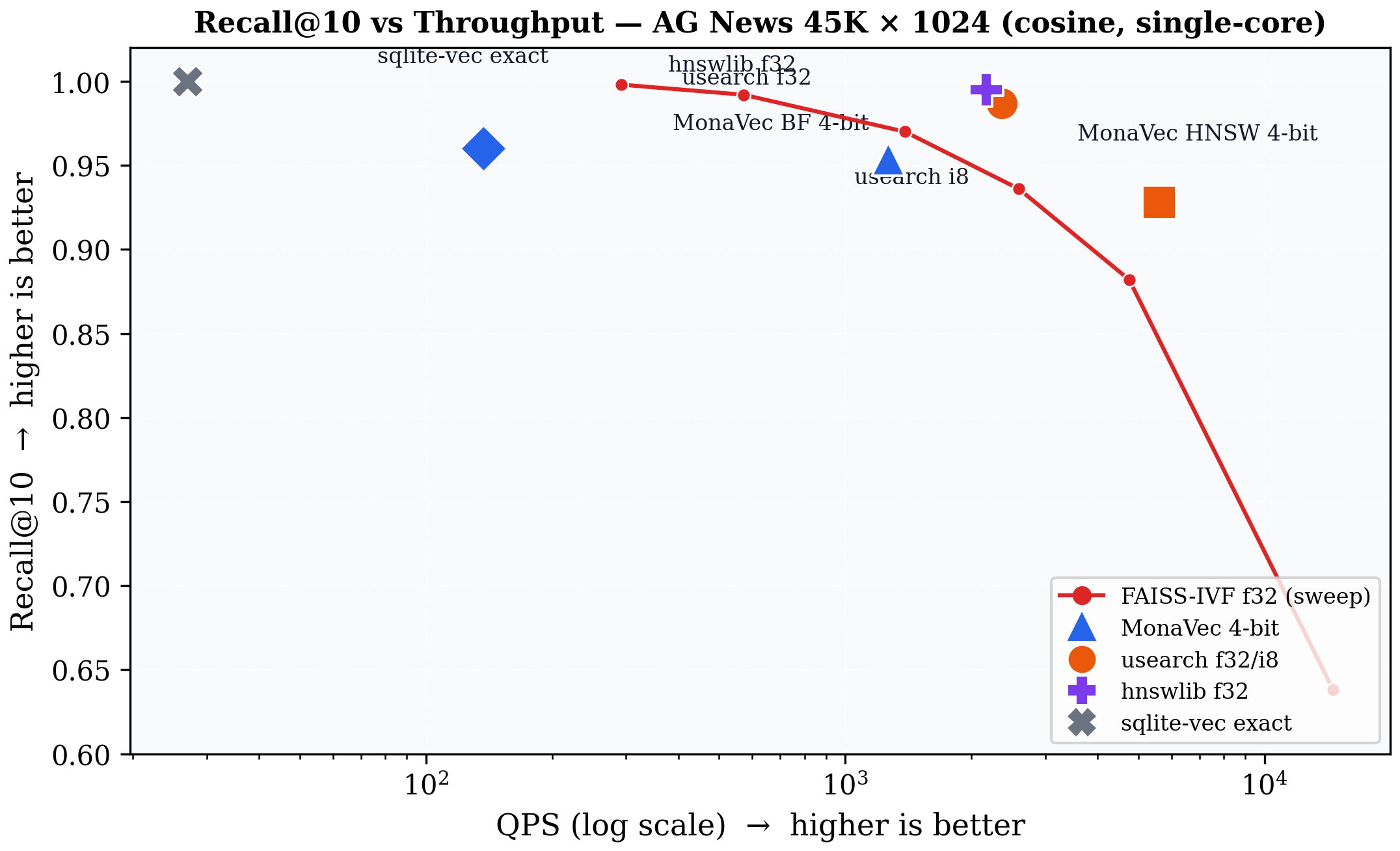}
    \caption{
        Recall@10 vs.\ throughput (log QPS) on AG News 45K\,$\times$\,1024-dim,
        all systems single-core on an Intel i7-13620H (AVX2) under identical
        conditions. FAISS-IVF is shown
        as an \texttt{nprobe} sweep. \monavec{} leads the high-recall region;
        mature SIMD-optimized libraries lead the high-throughput region.
    }
    \label{fig:recall_qps}
\end{figure}

Figure~\ref{fig:recall_qps} places all evaluated systems on a single
recall--throughput plane (AG News, single-core, log QPS). The picture is
honest: \monavec{} occupies the \emph{high-recall} region, while usearch,
hnswlib, and current FAISS-IVF occupy the \emph{high-throughput} region.
\monavec{} HNSW sits above the FAISS-IVF sweep on recall at comparable
operating points but to its left on QPS; the uncompressed float32 graph
indexes lead on both recall and speed at the cost of $8\times$ memory.
\monavec{}'s distinct value is delivering near-exact recall at 4-bit memory
with full determinism—not winning the raw-throughput corner.

\section{Discussion}

This section examines the design boundaries of \monavec{}: which established
techniques we deliberately exclude from the training-free core (Product
Quantization, int8 query quantization), how the system relates to concurrent
rotation-based quantization work, the lessons that shaped the implementation, and
what the current evaluation does and does not yet cover.

\subsection{Why Not Product Quantization?}
\label{sec:why_not_pq}

Product Quantization (PQ)~\cite{jegou2011pq}—together with optimized variants such
as OPQ~\cite{ge2013opq} and anisotropic vector quantization~\cite{guo2020scann}—achieves
higher recall than scalar quantization on pixel data (fashion-mnist $\sim$0.85+ vs 0.62)
by jointly encoding groups of dimensions with a data-specific codebook.
\monavec{} deliberately excludes PQ from the core for three reasons:

\begin{enumerate}[leftmargin=1.5em, itemsep=2pt]
    \item \textbf{Training requirement.} PQ requires k-means training
          over the indexed corpus. This violates the zero-training constraint
          and prevents cold-start indexing (first document, first query).

    \item \textbf{Stored codebook.} PQ codebooks (256 centroids per sub-space,
          $\sim$50 sub-spaces for 1536-dim) add $\sim$50KB to every index file.
          For edge devices handling many small indexes, this overhead is significant.

    \item \textbf{Scope.} MonaVec's primary target is semantic embeddings
          (cosine similarity, unit-norm inputs), where scalar quantization
          already achieves 0.960 recall—PQ yields marginal gains at substantial
          complexity cost.
\end{enumerate}

PQ is planned as an optional feature flag (\texttt{features = ["pq"]}) for
users with raw-magnitude L2 data who can afford a training pass.
The core will remain training-free.

\subsection{Why Not Int8 Query Quantization?}
\label{sec:why-not-int8}

usearch's i8 mode is its fastest configuration (5{,}726 QPS on AG News) because
it quantizes \emph{both} sides to 8-bit integers and scores with a single
integer dot product (VNNI/\texttt{simsimd})—no nibble unpack, no float lookup.
It would be tempting to match this by quantizing \monavec{}'s query to int8 as
well. We deliberately do not, for two reasons aligned with the system's identity:

\begin{enumerate}[leftmargin=1.5em, itemsep=2pt]
    \item \textbf{Asymmetric scoring protects recall.} \monavec{} keeps the
          query in full float32 and only the database side is quantized
          (\S\ref{sec:asymmetric}). Quantization error then enters the dot
          product on \emph{one} side only. Symmetric int8 scoring would add error
          on both sides, eroding the recall lead that is our core advantage—on
          AG News we already beat usearch-i8 (0.960 vs 0.928) precisely because
          our query retains full precision at 4-bit storage.
    \item \textbf{Determinism.} An int8 query path introduces a second
          quantization step whose rounding interacts with the SIMD reduction
          order, widening the cross-platform divergence surface we work to keep
          closed (Section~2). The float32 query keeps the query-side computation
          exact.
\end{enumerate}

In short, usearch reaches its peak throughput by trading away precision symmetry;
that trade sells the very property—high recall at aggressive compression—that
distinguishes \monavec{}. We instead pursue throughput through latency-hiding in
the scoring kernel (\S\ref{sec:scoring-kernel}) and the fixed-baseline build
(\S3.7), which raise QPS \emph{without} touching query precision or determinism.

\subsection{Relation to TurboQuant}
\label{sec:turboquant}

TurboQuant~\cite{turboquant} (Zandieh et al., 2025; subsequently at ICLR 2026)
shares \monavec{}'s core quantization principle: randomly rotate inputs to
concentrate the coordinate distribution, then exploit high-dimensional
near-independence to apply an optimal \emph{per-coordinate scalar} quantizer.
\monavec{} reaches a similar data-oblivious recipe—RHDH rotation followed by
Lloyd-Max scalar quantization—from a different modeling assumption: TurboQuant
quantizes against the \emph{exact} concentrated Beta marginal of a rotated unit
vector, whereas \monavec{} uses the $\mathcal{N}(0,1)$ Gaussian limit of that
marginal (the CLT regime of the Walsh--Hadamard transform), which is accurate at
the embedding dimensionalities we target and admits a single precomputed table.
TurboQuant develops the idea as a general-purpose quantization \emph{algorithm}
with near-optimal distortion-rate guarantees and applications such as KV-cache
compression.

\monavec{}'s contribution is orthogonal and complementary: it is not a new
quantizer but a complete \emph{embedded vector-search system} in which a
data-oblivious quantizer is co-designed with the index (BruteForce/IvfFlat/HNSW),
the asymmetric scoring path, portable determinism, single-file persistence, and a
service layer. The quantization stage could in principle be swapped for
TurboQuant's; the system contributions—metric-aware graph construction, auto-M,
zero-copy ingestion, the deterministic single-file format, and the
embedded/offline deployment model—are where \monavec{} stands apart. In short,
TurboQuant answers ``how well can a data-oblivious quantizer compress?''; \monavec{}
answers ``how does such a quantizer become a deterministic, dependency-free search
engine on the edge?''

\subsection{Lessons Learned}

\paragraph{Graph topology and metric must be aligned.}
Using dot product scoring for HNSW graph construction when the search metric
is L2 was a subtle but impactful bug. Graph traversal during construction
navigates toward ``similar by dot product'' neighbors; during search, it ranks
by L2. The resulting graph topology is mismatched, and greedy search
consistently fails to find the true nearest neighbors. The fix—using
$\langle q, v \rangle - \frac{1}{2}\norm{v}^2$ during construction for L2—is
simple but non-obvious. This underscores that quantized ANN systems must
apply metric awareness at every stage: encoding, indexing, traversal,
and scoring.

\paragraph{Global standardization preserves ordering; per-dimension does not.}
Our initial L2 fix used per-dimension whitening (Mahalanobis standardization),
which achieves better quantizer fit but changes the search metric.
Recall@10 with per-dimension whitening (0.53) is worse than global scalar
standardization (0.62) on fashion-mnist, confirming that metric preservation
dominates quantizer optimality for nearest-neighbor search.

\paragraph{M must scale with N for HNSW.}
At N=1.18M, M=32 is empirically insufficient. The graph diameter argument
provides the theoretical explanation: fixed M means O($\log N$) layers
but also growing ``local minimum'' probability as the search space expands.
M=64 does not eliminate this problem—it reduces it.
For very large N ($\gg$10M), further M increases may be warranted;
we leave this to future work.

\subsection{Limitations and Roadmap}
\label{sec:limitations}

\begin{itemize}[itemsep=4pt]
    \item \textbf{Dot/L2 quantization on unnormalized vectors.}
          For heavily unnormalized Dot inputs, Lloyd-Max tables remain
          suboptimal. A \texttt{fit()} analog for Dot metric is planned.

    \item \textbf{Product Quantization.}
          PQ (\texttt{features = ["pq"]}) will be added for L2 use cases
          where the training pass is acceptable. This keeps the core
          training-free while enabling FAISS-class recall for raw-magnitude data.

    \item \textbf{2-bit SIMD.}
          The 2-bit scoring path uses scalar fallback; AVX2/NEON 2-bit
          kernels are planned.

    \item \textbf{Optimized NEON 4-bit kernel.}
          The ARM 4-bit scoring path currently runs the scalar reference after a
          correctness fix (\S\ref{sec:edge}). A correct NEON table-lookup kernel
          (\texttt{vqtbl4q\_u8}) would restore SIMD throughput on edge devices
          without sacrificing correctness or determinism.

    \item \textbf{Mixed-precision on real workloads.}
          The mixed-precision recall--compression advantage
          (Figure~\ref{fig:mixed}) is currently demonstrated on synthetic
          Gaussian data only. Evaluating it on the semantic and pixel
          benchmarks, and against external baselines, remains future work.

    \item \textbf{Evaluation scope.}
          Our experiments cover the quantization pipeline, the three index
          backends, and their recall/throughput/memory behaviour. The remaining
          system features described in Section~\ref{sec:competitors} and
          Section~3---hybrid sparse-dense retrieval (BM25 + RRF), the pre-filter
          allowlist, and identity-based multi-tenancy---are presented as design
          contributions; their empirical evaluation is future work.

    \item \textbf{Quantization-method baselines.}
          We benchmark against packaged embedded libraries (usearch, hnswlib,
          sqlite-vec, FAISS-IVF). A direct comparison against quantization
          \emph{techniques}---RaBitQ~\cite{gao2024rabitq} and the FAISS scalar
          quantizer at a matched 4-bit width---would isolate the quantizer from
          the surrounding system and is planned.

    \item \textbf{Reported throughput variance.}
          QPS is reported as a point estimate (best timed pass after warm-up)
          rather than a distribution over repeated runs. A mean\,$\pm$\,std
          characterization across executions and machines is future work; because
          the pipeline is deterministic, recall itself is exactly reproducible, so
          the variance of interest is confined to timing.

    \item \textbf{Single-threaded search.}
          Designed for single-thread embedded use. Application-level
          parallelism across queries can be used where throughput is critical.

    \item \textbf{ColPali / multi-vector late interaction.}
          Document-level embeddings via ColBERT~\cite{khattab2020colbert} /
          ColPali~\cite{faysse2024colpali} are planned
          (\texttt{features = ["colpali"]}), targeting visual document
          understanding without a server.
\end{itemize}

\section{Conclusion}

We presented \monavec{}, a training-free embedded vector search kernel for
edge and offline AI systems. The system is built on three core insights:
(1) \rhdh{} rotation produces $\mathcal{N}(0,1)$-distributed coordinates
regardless of input distribution, enabling precomputed Lloyd-Max quantization
tables without any data pass;
(2) metric awareness must be applied at every pipeline stage—encoding,
graph construction, traversal, and scoring—not only at query time;
(3) the deployment model of SQLite (one file, one call, runs anywhere) is
the right model for vector search on the edge.

On semantic embeddings, \monavec{} 4-bit quantization leads float32 FAISS-IVF
on recall (0.954--0.960 vs 0.936) at $8\times$ smaller memory; FAISS retains a
throughput advantage that reflects mature hand-tuned SIMD rather than an
architectural ceiling on our side. On raw-magnitude L2 data, global scalar
standardization recovers 52\% of the recall lost without preprocessing,
while correctly preserving Euclidean distance ordering. At 1M+ scale,
auto-M selection addresses graph diameter growth with a simple and
empirically validated policy.

The system ships as a Rust crate, a Python package, and a complete service
stack including REST API, admin UI, hybrid sparse-dense retrieval, and
identity-based multi-tenancy—all running offline, with no external
dependencies. We believe \monavec{} fills a genuine gap in the edge AI
ecosystem: as on-device LLMs become commodity, the retrieval layer should
meet the same deployment constraints.

Head-to-head against mature embedded libraries, \monavec{} today leads on recall
at 4-bit compression—beating FAISS-IVF and 8-bit usearch on AG News, and every
graph index we evaluated on million-scale glove-100—and is the only system offering portable,
byte-identical determinism, while trailing 2--15$\times$ on raw throughput. We are
deliberate about this profile: the recall and determinism advantages are
mathematical and durable, while the throughput deficit is an engineering gap we aim
to narrow through SIMD latency-hiding, graph-memory compression, and wider-vector
kernels—\emph{each chosen to preserve query precision and determinism rather than
trade them for speed}. \monavec{} is not trying to become a faster HNSW library;
it is establishing a different point in the design space—maximal correctness and
reproducibility per byte—and engineering its way toward making throughput a
non-concern for the edge and offline deployments it targets.

\section*{Acknowledgements}

The author thanks the maintainers of the open-source projects this work
benchmarks against---FAISS, usearch, hnswlib, and sqlite-vec---whose public
implementations made the comparative evaluation possible, and the colleagues who
reviewed earlier drafts and whose feedback materially improved the manuscript.
The author is grateful to Prof.~Dr.~Şakir Taşdemir and Dr.~Merve Ayyüce
Kızrak for their valuable contributions to the development of this work. Finally,
the author thanks his beloved wife for her constant support, and his children
Utkan Uras Yenen and Arden Aras Yenen for their patience during the long and
demanding effort this work required.

\section*{Declaration on the Use of AI Tools}

The author used AI-based assistants (large language models) to support manuscript
drafting, code implementation, and editing. All research contributions—the system
design, quantization pipeline, index backends, experimental methodology,
benchmarks, and analysis—were conceived, executed, and verified by the author.
All bibliographic references were checked against primary sources, and all
empirical results were produced and validated on the hardware described in
Section~\ref{sec:competitors} and the Evaluation section. The author takes full
responsibility for the content of this paper.

\bibliographystyle{plain}

\end{document}